\documentclass[twocolumn,english,prb,showpacs]{revtex4-1}
\usepackage[colorlinks=true,urlcolor=blue,citecolor=blue,linkcolor=blue]{hyperref} 
\usepackage[T1]{fontenc} 
\usepackage[latin9]{inputenc} 
\usepackage{amssymb} 
\usepackage{amsmath,color} 
\usepackage{graphicx} 
\usepackage{babel} 
\usepackage{esint} 
\usepackage{float} 
\usepackage{indentfirst} 
\usepackage{braket} 
\usepackage{mathrsfs}
\usepackage{blkarray}

\newcommand{\Tr}{{\rm Tr}}

\newcommand{\Eq}[1]{Eq.~(\ref{#1})}
\newcommand{\Fig}[1]{Fig.~\ref{#1}}
\newcommand{\Ref}[1]{Ref.~\onlinecite{#1}}
\newcommand{\Refs}[1]{Refs.~\onlinecite{#1}}

\newcommand{\red}[1]{\textcolor{red}{\bf #1}}

\begin{document}

\title{Quantum Monte Carlo study of mass-imbalanced Hubbard models}

\author{Ye-Hua Liu}
\author{Lei Wang}
\affiliation{Theoretische Physik, ETH Zurich, 8093 Zurich, Switzerland} 

\begin{abstract} 
Building on recent solutions of the fermion sign problem for specific models we present two continuous-time quantum Monte Carlo methods for efficient simulation of mass-imbalanced Hubbard models on bipartite lattices at half-filling. For both methods we present the solutions to the fermion sign problem and the algorithms to achieve efficient simulations. As applications, we calculate the dependence of the spin correlation on the mass imbalance in a one-dimensional lattice and study the thermal and quantum phase transitions to an antiferromagnetic Ising long-range ordered state in two dimensions. These results offer unbiased predictions for experiments on ultracold atoms and bridge known exact solutions of Falicov-Kimball model and previous studies of the $SU(2)$-symmetric Hubbard model. 
\end{abstract}


\maketitle



\section{Introduction}
Recently, progress has been made in solutions of the sign problem for fermionic models with specific symmetries.\cite{Huffman:2014fj, Li:2015jf, Wang:2015vw} Combined with the development of efficient continuous-time quantum Monte Carlo (CT-QMC) approach for lattice fermions\cite{Iazzi:2014vv, Wang:2015tf} these advances enable the  unbiased simulation of low temperature phases of several fermionic models that were previously prohibitive, thereby quantitatively addressing long standing questions\cite{Scalapino:1984wz, Gubernatis:1985wo, Anonymous:xaWVK-gC} such as the fermionic quantum critical point of spinless fermions on the honeycomb lattice.\cite{Wang:2014iba, Wang:2015tf, Li:2015cwb}

In this paper we build on these conceptual breakthroughs and present two CT-QMC methods for efficient simulation of half-filled mass-imbalanced Hubbard models on bipartite lattices. Here, the term \emph{mass imbalance} refers to unequal hopping amplitudes for spin-up and spin-down fermions, {\it i.e.} we consider the Hamiltonian
\begin{align}
\hat{H} = 
&- \sum_{\sigma\in\left\{\uparrow,\downarrow\right\}}  t_{\sigma} \sum_{\braket{\mathbf{i,j}}} \left(\hat{c}_{\mathbf{i}\sigma}^{\dagger}\hat{c}_{\mathbf{j}\sigma}+\hat{c}_{\mathbf{j}\sigma}^{\dagger}\hat{c}_{\mathbf{i}\sigma}\right)\nonumber\\
&+U\sum_{\mathbf{i}}\left(\hat{n}_{\mathbf{i}\uparrow}-\frac{1}{2}\right)\left(\hat{n}_{\mathbf{i}\downarrow}-\frac{1}{2}\right),
\label{eq:Ham} 
\end{align}
where ${\hat c}_{\mathbf{i}\sigma}$ (${\hat c}^\dagger_{\mathbf{i}\sigma}$) is the fermion annihilation (creation) operator for site $\mathbf{i}$ and spin $\sigma$, $\hat{n}_{\mathbf{i}\sigma} = \hat{c}^\dagger_{\mathbf{i}\sigma} \hat{c}_{\mathbf{i}\sigma} $ is the fermion number operator, $t_{\uparrow(\downarrow)}$ is the hopping amplitude of the spin up (down) particles, $U>0$ denotes the on-site repulsive interaction between the two spin species. \footnote{At half-filling the physics of $U<0$ is simply related by a particle-hole transformation.} $\braket{\mathbf{i,j}}$ denotes two nearest neighbor sites belonging to different sublattices. When $t_{\uparrow}\neq t_{\downarrow}$ the $SU(2)$ symmetry in the spin space and the time-reversal symmetry are both broken. Such a Hubbard model with unequal hopping amplitudes can be readily implemented in an optical lattice by loading mixtures of ultracold fermionic atoms with different masses.\cite{PhysRevLett.100.010401,PhysRevLett.103.223203, PhysRevLett.104.053202, PhysRevLett.105.190401, PhysRevLett.106.115304, Kohstall:2013kg, Jag:2014gd} Furthermore, by using spin-dependent modulations, the group of T. Esslinger has recently realized this model in a one-dimensional optical lattice with a continuously tunable mass imbalance $t_{\downarrow}/t_{\uparrow}$.\cite{Jotzu:2015tq}

In the strong coupling limit $U\gg t_{\uparrow},t_{\downarrow}$, the low energy physics of the mass-imbalanced Hubbard model is captured by the following spin-$1/2$ XXZ model: 
\begin{equation}
\hat{H}_\mathrm{XXZ}=\sum_{\braket{\mathbf{i,j}}}\frac{4t_{\uparrow}t_{\downarrow}}{U}\left(\hat{S}_{\mathbf{i}}^{x}\hat{S}_{\mathbf{j}}^{x}+ \hat{S}^{y}_{\mathbf{i}}\hat{S}_{\mathbf{j}}^{y}\right) +\frac{2(t_{\uparrow}^{2}+t_{\downarrow}^{2})}{U}\hat{S}_{\mathbf{i}}^{z}\hat{S}^{z}_{\mathbf{j}}, 
\label{eq:xxz}
\end{equation}
where $\hat{S}^{\alpha}_\mathbf{i}=\frac{1}{2}\sum_{\mu\nu}\hat{c}_{\mathbf{i}\mu}^{\dagger} \sigma^{\alpha}_{\mu\nu} \hat{c}_{\mathbf{i}\nu}$ is the spin-$1/2$ operator and $\sigma^\alpha$ are the Pauli matrices. Since $2(t_{\uparrow}^{2}+t_{\downarrow}^{2})\ge 4t_{\uparrow}t_{\downarrow}$, the XXZ model has Ising anisotropy, and prefers longitudinal spin correlation $\braket{\hat{S}_{\mathbf{i}}^{z}\hat{S}_{\mathbf{j}}^{z}}$ than transverse correlations $\braket{ \hat{S}_{\mathbf{i}}^{x}\hat{S}_{\mathbf{j}}^{x}}$ or $\braket{ \hat{S}_{\mathbf{i}}^{y}\hat{S}_{\mathbf{j}}^{y}}$. 
On the other hand, model~(\ref{eq:Ham}) reduces to the Falicov-Kimball model when $t_{\downarrow}=0$,\cite{Falicov:1969ua} which describes a mixture of localized heavy particles and itinerant light fermions interacting through onsite repulsions. This limit allows various exact analytical and numerical studies.\cite{brandt1986exact, PhysRevLett.88.106401, 2003RvMP...75.1333F,Maska:2006id, Zonda:2010io} In particular, the low temperature phase on bipartite lattices was proven to be a staggered density-wave state of both species for arbitrary repulsive interactions.\cite{Kennedy:1986wf} In agreement with the strong coupling analysis, this state possesses an antiferromagnetic Ising long range order. 

Various aspects of the mass-imbalanced Hubbard model for general finite hoppings $t_{\uparrow}\neq t_{\downarrow}$ have been the subjects of intensive research.\cite{PhysRevB.52.13910, PhysRevB.76.104517, PhysRevA.85.013606, PhysRevLett.109.065301, PhysRevA.90.023605, Winograd:2011en, Winograd:2012ie, PhysRevB.77.085110, Roscher:2014hb, Braun:2015du, Gezerlis:2009fb, Kroiss:2015dy} Bosonic versions of the model~(\ref{eq:Ham}) were studied in \Refs{Soyler:2009eya, PhysRevA.81.053622}. 
However, the traditional determinantal QMC\cite{Blankenbecler:1981vj} method faces a severe sign problem even at half-filling on bipartite lattices when applied to the model (\ref{eq:Ham}), because the breaking of  time-reversal symmetry makes it difficult to relate the determinants of spin up and down components.\cite{Wu:2005im} As a consequence, despite its simple form and fundamental importance, an unbiased study of model (\ref{eq:Ham}) at half-filling in more than  one dimension has not yet been performed.

In this paper we first present in detail two CT-QMC methods that solve the model~(\ref{eq:Ham}) by using recent advances regarding the fermion sign problem.\cite{Huffman:2014fj, Li:2015jf, Wang:2015vw} One  method is based on the continuous-time interaction-expansion (CT-INT) approach\cite{Rubtsov:2005iw} whose sign problem is solved based on~\Ref{Huffman:2014fj}. However a naive CT-INT simulation of model (\ref{eq:Ham}) suffers from low acceptance rate and also difficulties in measuring two-particle correlations. We present solutions to these problems making use of correlated double-vertex updates and shift moves in the Monte Carlo simulation. The second method is a recent LCT-AUX approach,\cite{Iazzi:2014vv} which has $\mathcal{O}(\beta U N^3)$ scaling with respect to the inverse temperature $\beta$ and the lattice size $N$. The sign problem of LCT-AUX approach is solved thanks to a recent recognition of the Lie group and Lie algebra structures of the determinantal QMC approaches.\cite{Wang:2015vw} 

\begin{figure}[t]
\includegraphics[width=8.8cm]{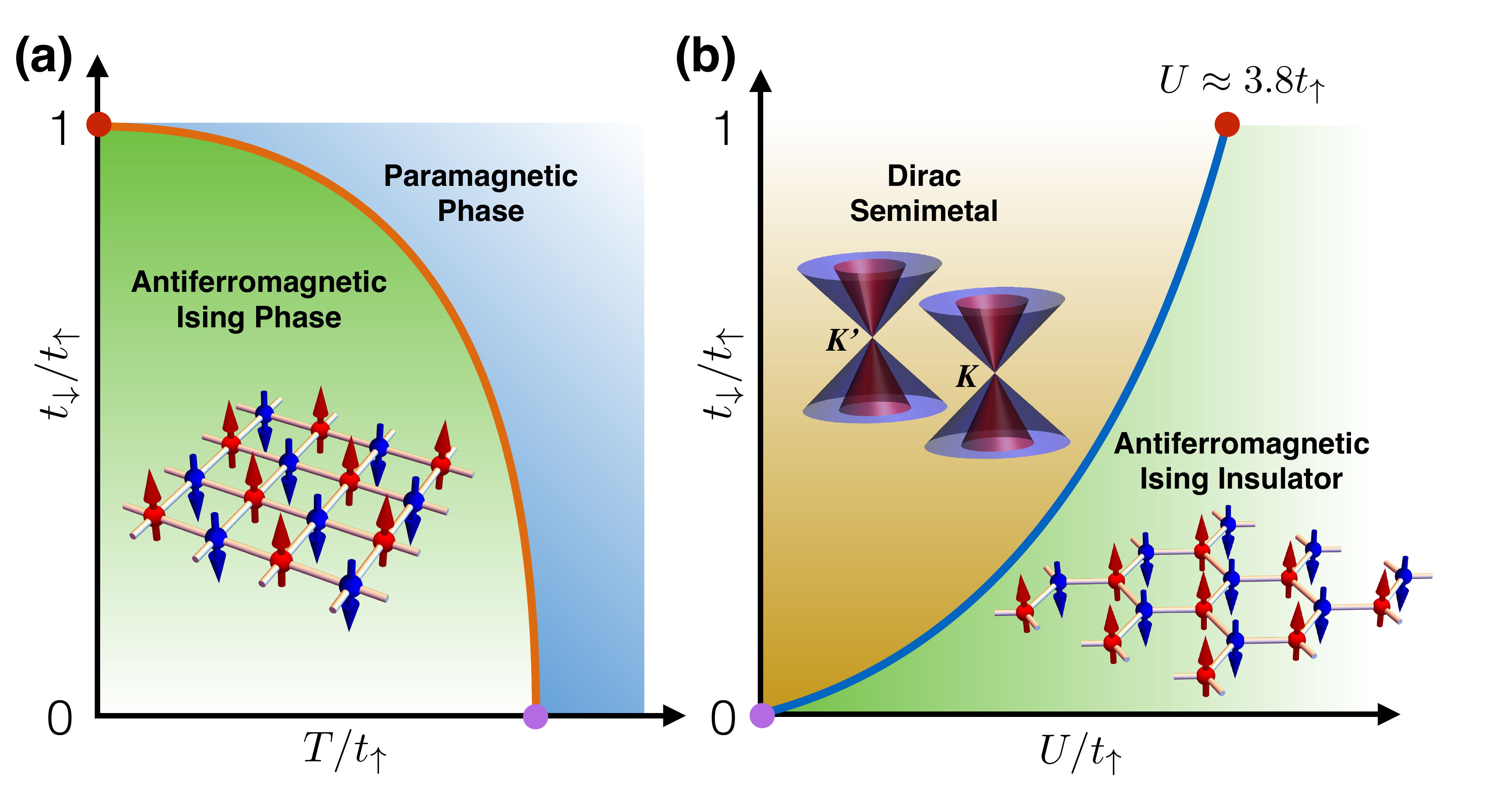}
\caption{Schematic plots of (a) the finite-temperature phase diagram on the square lattice and (b) the ground-state phase digram on the honeycomb lattice. 
The solid line in (a) is the phase boundary of a thermal phase transition while the solid line in (b) the phase boundary of a quantum phase transition. The red and purple dots indicate known critical points of the $SU(2)$ symmetric Hubbard model ($t_{\uparrow}=t_{\downarrow}$) and the Falicov-Kimball model ($t_{\downarrow}=0$) respectively. 
\label{fig:phasediag}}
\end{figure}

We then report results for magnetic properties and phase transitions of the mass-imbalanced Hubbard model on various lattices, as summarized in Fig.~\ref{fig:phasediag}. We provide quantitative predictions of the nearest-neighbor spin-spin correlations in a one-dimensional  lattice and the critical temperature to the antiferromagnetic long-range ordered state on a square lattice. Furthermore, we address the quantum phase transition of the Dirac semimetal phase to an antiferromagnetic Ising phase on the honeycomb lattice. All these predictions are closely relevant to current efforts in ultracold fermion experiments and can be verified through experimental observations.\cite{Greif:2013kb, Jotzu:2015tq}

\section{Methods \label{sec:methods}}

We start by rewriting Eq. (\ref{eq:Ham}) as $\hat{H}=\hat{H}_{0}+\sum_{\mathbf{i}}\hat{v}_\mathbf{i}$ with 
\begin{eqnarray}
\hat{H}_{0}& = & \sum_{\sigma}\sum_{\mathbf{ij}} \hat{c}_{\mathbf{i}\sigma}^{\dagger}K_{\mathbf{ij}}^{\sigma}\hat{c}_{\mathbf{j}\sigma}, \label{eq:H0} \\
\hat{v}_{\mathbf{i}}& = &U\left(\hat{n}_{\mathbf{i}\uparrow}\hat{n}_{\mathbf{i}\downarrow}-\frac{\hat{n}_{\mathbf{i}\uparrow}+\hat{n}_{\mathbf{i}\downarrow}}{2}\right)-\Gamma. \label{eq:v} 
\end{eqnarray}
The hopping matrix $K_{\mathbf{ij}}^{\sigma}=-t_{\sigma}$ when the sites $\mathbf{i}$, $\mathbf{j}$ are nearest neighbors, and zero otherwise. A constant shift $\Gamma$ in the local interaction term is introduced for later convenience. We then perform an interaction expansion for the partition function, 
\begin{eqnarray}
Z &=& \Tr \left(e^{-\beta \hat{H}}\right) 
= \sum_{k=0}^{\infty} \sum_{\mathbf{i}_{1},\ldots,\mathbf{i}_{k}} \int_{0}^{\beta}d \tau_{1} \cdots \int^{\beta}_{\tau_{k-1}} d\tau_{k} \nonumber \\ &\times& \Tr\left[e^{-(\beta-\tau_{k})\hat{H}_{0} } \left(-\hat{v}_{\mathbf{i}_{k}}\right) \cdots \left(-\hat{v}_{\mathbf{i}_{1}}\right) e^{-\tau_{1}\hat{H}_{0}} \right]  \nonumber
\\
& = & \sum_{k=0}^{\infty} \sum_{\mathcal{C}_{k}} w(\mathcal{C}_{k}).  \label{eq:expansion} 
\end{eqnarray}
In the last step we denote the summations and integrations of the first line abstractly as a sum over configurations $\mathcal{C}_{k}$. Monte Carlo methods sample these configurations stochastically according to the weight $w(\mathcal{C}_{k})$. Physical observables are evaluated as 
\begin{equation}
\braket{\hat{O}} = \frac{1}{Z}\Tr \left(e^{-\beta \hat{H}}\hat{O}\right) = \frac{1}{Z}\sum_{k=0}^{\infty} \sum_{\mathcal{C}_{k}}w(\mathcal{C}_{k}) \braket{\hat{O}}_{\mathcal{C}_{k},\tau},  
\end{equation} 
where 
\begin{equation}
\braket{\hat{O}}_{\mathcal{C}_{k},\tau}=\frac{\Tr\left[e^{-(\beta-\tau_{k})\hat{H}_{0} } \left(-\hat{v}_{\mathbf{i}_{k}}\right)\cdots \hat{O} \cdots \left(-\hat{v}_{\mathbf{i}_{1}}\right) e^{-\tau_{1}\hat{H}_{0}}\right]}{\Tr\left[e^{-(\beta-\tau_{k})\hat{H}_{0} } \left(-\hat{v}_{\mathbf{i}_{k}}\right) \cdots \left(-\hat{v}_{\mathbf{i}_{1}}\right) e^{-\tau_{1}\hat{H}_{0}} \right]}
\label{eq:estimator}
\end{equation}
is the expectation value of the observable inserted at imaginary time $\tau$ of a given configuration. In the following we denote \Eq{eq:estimator} as the estimator of the observable. The imaginary time index $\tau$ can take any value in $[0,\beta)$ because of translational invariance along  imaginary time. 

Key observables to identify the magnetic properties of model~(\ref{eq:Ham}) are staggered spin structure factors along various directions $S^{\alpha}_{\mathrm{AF}}=\frac{1}{N}\sum_\mathbf{ij}\eta_\mathbf{i}\eta_\mathbf{j} \braket{\hat{S}^{\alpha}_\mathbf{i}\hat{S}^{\alpha}_\mathbf{j}}$, where $\eta_{\mathbf{i}}=\pm 1$ is the parity of the lattice site.  Related to this quantity, one can further define 
\begin{eqnarray}
M_{2}& =& \left\langle\left(\frac{1}{N} \sum_\mathbf{i} \eta_\mathbf{i}\hat{S}^{z}_\mathbf{i}\right)^{2}\right\rangle=S_\mathrm{AF}^{z}/N, \label{eq:M2}  \\
M_{4}& =&\left\langle\left(\frac{1}{N} \sum_\mathbf{i} \eta_\mathbf{i}\hat{S}^{z}_\mathbf{i}\right)^{4}\right\rangle. \label{eq:M4}
\end{eqnarray}
They are the square and quartic power of the Ising order parameter respectively. Next, we present two sampling strategies and the corresponding measurement procedures for \Eq{eq:expansion}. The following two subsections can be read independently. 
 
\subsection{Continuous-time interaction expansion algorithm (CT-INT) \label{sec:CTINT}} 
In the CT-INT approach we choose a special value for the constant shift $\Gamma = -U/4$ in \Eq{eq:v} so that $\hat{v}_\mathbf{i}=U\left(\hat{n}_{\mathbf{i}\uparrow}-\frac{1}{2}\right)\left(\hat{n}_{\mathbf{i}\downarrow}-\frac{1}{2}\right)$. 
Using Wick's theorem in \Eq{eq:expansion}, all possible contractions add up to a determinant for each spin component\cite{Rubtsov:2005iw}
\begin{align}
Z= 
Z_{0}\sum_{k=0}^{\infty} \sum_{\mathcal{C}_{k}} 
\left(-U\right)^{k}\prod_{\sigma} \det \left(G^{\sigma}\right), 
\label{eq:ctint} 
\end{align}
where $Z_{0}=\Tr(e^{-\beta\hat{H}_{0}})$ is the noninteracting partition function and the set $\mathcal{C}_{k}=\left\{ \left(\mathbf{i}_{1},\tau_{1}\right),\ldots,\left(\mathbf{i}_{k},\tau_{k}\right)\right\} $ denotes a  configuration with $k$ vertices. 
$G^{\sigma}$ is a $k\times k$ matrix with matrix elements 
\begin{align}
G^{\sigma}_{pq} = \mathcal{G}_{\mathbf{i}_{p}\mathbf{i}_{q}}^{\sigma}\left(\tau_{p}-\tau_{q}\right)-\frac{1}{2}\delta_{pq}, 
\label{eq:G}
\end{align}
where $\mathcal{G}_{\mathbf{ij}}^{\sigma}\left(\tau \right) = Z_{0}^{-1}\Tr\left[ e^{-\beta\hat{H}_{0}}  \mathcal{T}\hat{c}_{\mathbf{i}\sigma}\left(\tau\right)\hat{c}_{\mathbf{j}\sigma}^{\dagger} \right] $ is the noninteracting Green's function and $\mathcal{T}$ is the time-ordering operator. Because of the particle-hole symmetry in model (\ref{eq:Ham}), $\mathcal{G}^{\sigma}_{\mathbf{ii}}(0^{+})\equiv 1/2$, thus the diagonal elements of $G^{\sigma}$ actually all vanish. 

\subsubsection{Absence of the sign problem \label{sec:sign problem}}


It is well-understood that model (\ref{eq:Ham}) is free from sign problem when $t_\uparrow=t_\downarrow$ because the determinants for the two spin components are then identical \emph{and} they are nonzero only for even expansion orders.\cite{Rubtsov:2005iw} However the absence of sign problem for general unequal hoppings was not appreciated until a recent discovery of~\Ref{Huffman:2014fj}. In below we summarize a simplified proof following Ref.~\onlinecite{Wang:2014iba}. 
On bipartite lattices 
one has 
\begin{equation}
({G^{\sigma}})^{T}=-D G^{\sigma}D, \label{eq:Gsymmetry}
\end{equation}
where $D$ is a diagonal matrix with $D_{pp}=\eta_{\mathbf{i}_{p}}$ consisting of parities of the lattice sites. The Monte Carlo weight in \Eq{eq:ctint} can then be written as 
\begin{eqnarray}
w(\mathcal{C}_{k})/Z_{0} & = & \left(-U\right)^{k} \prod_{\sigma} \det (-DG^{\sigma}D) \nonumber\\ &= & \left(-U\right)^{k} \prod_{\sigma} \det (DG^{\sigma}) \ge 0, 
\label{eq:ctint-sign}
\end{eqnarray}
where the second equality follows from the fact that the two spin species have the \emph{same} $D$ matrix even though their hopping amplitudes are not the same. Finally, since the matrix $DG^{\sigma}$ is real and antisymmetric following \Eq{eq:Gsymmetry}, its determinant is zero for odd expansion order $k$ and nonnegative for even $k$. Because each individual factor of \Eq{eq:ctint-sign} is nonnegative, there is no sign problem for neither repulsive nor attractive interaction $U$. 

\subsubsection{Monte Carlo updates \label{sec:pair_update}}
Usual CT-QMC updates consist of random insertion and removal of vertices in \Eq{eq:ctint}.\cite{Gull:2011jd} Here, because of the vanishing of Monte Carlo weights for odd expansion orders in \Eq{eq:ctint-sign}, one needs to insert or remove \emph{at least} two vertices together\cite{Rubtsov:2005iw, Kozik:2013ji, Anonymous:2014by} to ensure ergodicity of the sampling.
\footnote{When $t_{\uparrow}=t_{\downarrow}$, one could use a shift tuned slightly away from $\Gamma=-U/4$ to gain finite weight for odd expansion orders,\cite{Rubtsov:2005iw, Assaad:2007be} thus avoiding the inconvenience of correlated double-vertex updates. However, for general asymmetric $t_{\uparrow}\neq t_{\downarrow}$ case this leads to sign problem in the CT-INT simulation.} However, we observe these type of updates suffer from low acceptance rate if the two vertices are chosen independently 
since the preferred configurations have correlations between the vertices.\cite{Wang:2015bva} 

\begin{figure}[t]
\includegraphics[width=8cm]{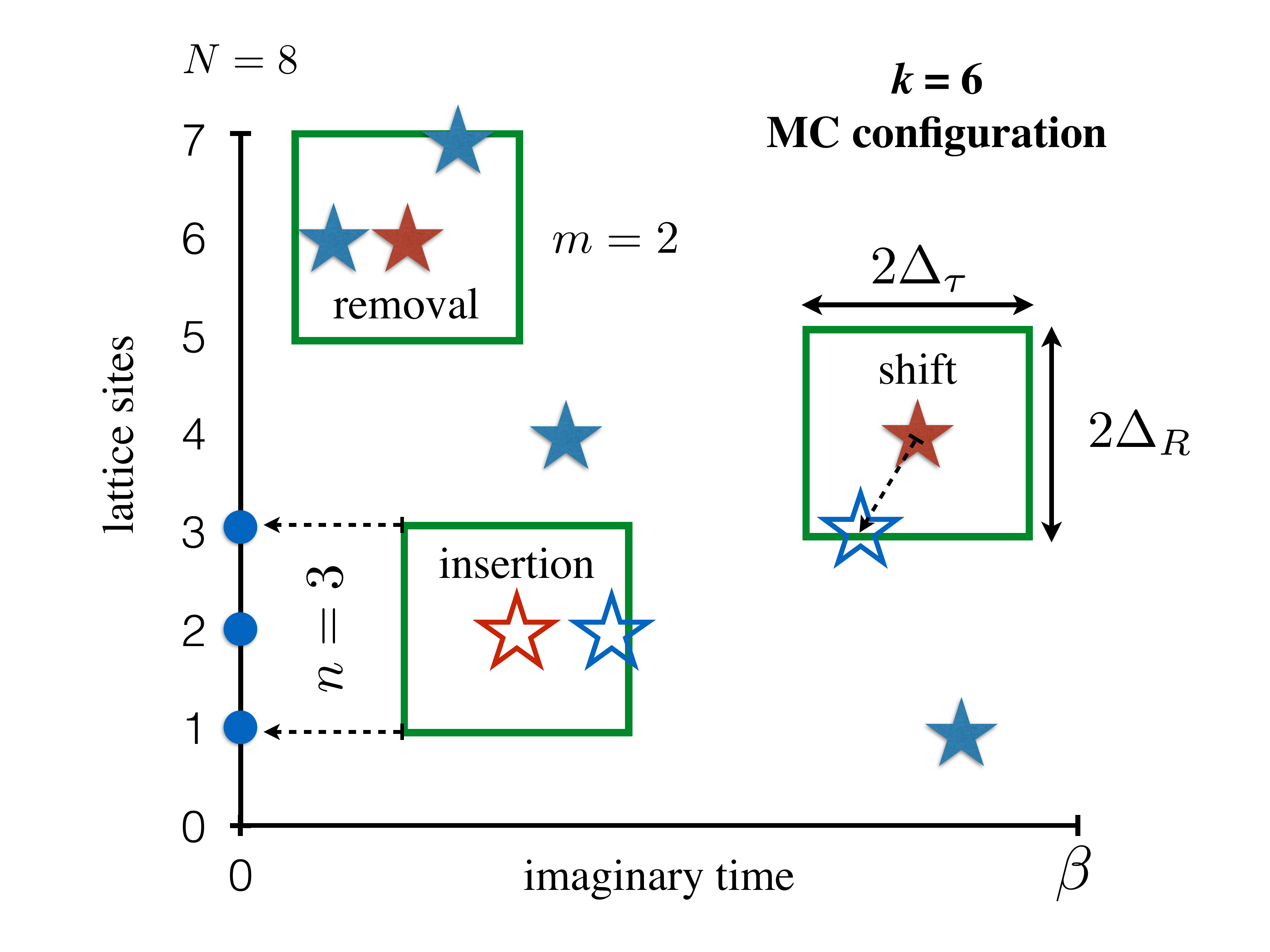}
\caption{Schematics of correlated double-vertex updates in CT-INT, shown for a one-dimensional lattice of length $N=8$. 
Vertices in the current Monte Carlo configuration are represented by solid stars (there are $k=6$ in the figure). While the candidates for the new configuration are marked as empty stars. We use red color to denote the first chosen vertex in the correlated double-vertex updates, and blue the second. The green box around each red vertex includes all nearby vertices. For insertion, the green box, when projected to the lattice-site axis, includes $n=3$ candidate sites for the second vertex; for removal, the green box encloses additional $m=2$ vertices other than the first chosen (red) one; for shift, a random position in the green box is proposed.
\label{fig:pair_update}}
\end{figure}

To overcome this low acceptance rate problem, we adopt a \emph{correlated} double-vertex update scheme illustrated in Fig.~\ref{fig:pair_update}. Only pairs of vertices, say $\left(\mathbf{i}_{p},\tau_{p}\right)$ and $\left(\mathbf{i}_{q},\tau_{q}\right)$, that are ``close'' in space time are inserted or removed. We define ``close'' to mean $\left|\mathbf{i}_{p}-\mathbf{i}_{q}\right|\le\Delta_{R}$ and  $\left|\tau_{p}-\tau_{q}\right|<\Delta_{\tau}$ for some predefined cutoffs $\Delta_{R}$ and $\Delta_{\tau}$. The definition of distances in space and time both take into account the periodicity of the lattice and the imaginary-time axis. In particular, the real-space distance $\left|\mathbf{i}_{p}-\mathbf{i}_{q}\right|$ is defined as the graph distance between the two sites $\mathbf{i}_{p}$ and $\mathbf{i}_{q}$. Our benchmark shows the correlated double-vertex insertion/removal updates greatly enhance the acceptance probability. 
To further reduce the autocorrelation time, we complement these moves with a shift update, where a vertex is shifted to a nearby location in space time, as shown in Fig.~\ref{fig:pair_update}. 

Starting from a configuration $\mathcal{C}_k$, for vertex insertion, the first vertex $\left(\mathbf{i}_{p},\tau_{p}\right)$ is picked uniformly at $\tau_{p}\in\left[0,\beta\right)$ and $\mathbf{i}_{p}\in\left[0,N\right)$. For the second vertex $\left(\mathbf{i}_{q},\tau_{q}\right)$, $\tau_{q}$ is picked uniformly randomly within a window of width $\Delta_{\tau}$ centered at $\tau_{p}$, and $\mathbf{i}_{q}$ randomly from sites within  distance $\Delta_{R}$ from site $\mathbf{i}_{p}$ (including $\mathbf{i}_{p}$ itself). The number of possible sites for $\mathbf{i}_{q}$ is denoted as $n$.\footnote{$n$ is a constant that only depends on the lattice geometry and the value of $\Delta_R$.} For the reverse move, one  picks the first vertex randomly from the existing $k+2$ vertices, and selects the second one randomly from all the other existing vertices that are ``close'' to the first one, with $m$ possible candidates, see \Fig{fig:pair_update}. Taking into account these proposal probabilities, the acceptance rate for insertion $p(\mathcal{C}_{k}\rightarrow \mathcal{C}_{k+2})=\min\left(1,r\right)$ and removal $p(\mathcal{C}_{k+2}\rightarrow \mathcal{C}_{k})=\min\left(1,r^{-1}\right)$, with the ratio  
\begin{align}
r =  \frac{U^{2}\beta N\Delta_{\tau} n }{\left(k+2\right) m } \prod_{\sigma} \frac{\det G^{\sigma}(\mathcal{C}_{k+2}) }{\det G^{\sigma}(\mathcal{C}_k) }, 
\end{align}
where $\mathcal{C}_{k+2}=\mathcal{C}_{k}\cup(\mathbf{i}_{p}, \tau_{p})\cup(\mathbf{i}_{q}, \tau_{q})$ and 
$G^{\sigma}(\mathcal{C}_{k+2})$ is different from $G^{\sigma}(\mathcal{C}_{k})$ by \emph{adding} two rows and two columns. We keep track of the inverse of the matrix $G^{\sigma}$ in the simulation, thus the calculation of the determinant ratio and the update can be done with $\mathcal{O}(k^{2})$ operations using the standard fast-update approach.\cite{Rubtsov:2005iw, Gull:2011jd}
   
For the shift update one 
 randomly selects an existing vertex $\left(\mathbf{i},\tau\right)$ from a configuration $\mathcal{C}_{k}$ and proposes to shift it to a new position $\left(\mathbf{i}',\tau'\right)$.\footnote{Long distance shift will be rejected with high probability. We therefore also impose the ``closeness'' condition and only shift the vertices within the cutoffs $\Delta_{\tau}$ and $\Delta_{R}$.} 
The proposal probabilities for forward and backward shifts equal and the acceptance probability is simply the ratio of the Monte Carlo weights
\begin{equation}
p(\mathcal{C}_{k}\rightarrow \mathcal{C}_{k}^{\prime})=\min\left(1, \prod_{\sigma} \frac{\det \left(G^{\sigma}(\mathcal{C}^{\prime}_{k}) \right)}{\det \left(G^{\sigma}(\mathcal{C}_{k})\right) }\right), \label{eq:shiftratio}
\end{equation}
where $\mathcal{C}^{\prime}_{k}=\mathcal{C}_{k}\setminus(\mathbf{i},\tau)\cup(\mathbf{i}^{\prime},\tau^{\prime})$ and $G^{\sigma}(\mathcal{C}^{\prime}_{k})$ is different from $G^{\sigma}(\mathcal{C}_{k})$ by \emph{changing} one row and one column. As the notation suggests, one way to calculate the ratio in \Eq{eq:shiftratio} is to first remove the vertex $(\mathbf{i}, \tau)$ and then insert back $(\mathbf{i}^{\prime}, \tau^{\prime})$, in which an intermediate configuration with $k-1$ vertices is reached, i.e. 
\begin{equation}
\frac{\det \left(G^{\sigma}(\mathcal{C}^{\prime}_{k}) \right)}{\det \left(G^{\sigma}(\mathcal{C}_{k})\right) }=\frac{\det \left(G^{\sigma}(\mathcal{C}^{\prime}_{k}) \right)}{\det \left(G^{\sigma}(\mathcal{C}_{k}\setminus(\mathbf{i},\tau))\right) } \frac{\det \left(G^{\sigma}(\mathcal{C}_{k}\setminus(\mathbf{i},\tau)) \right)}{\det \left(G^{\sigma}(\mathcal{C}_{k})\right) }.
\label{eq:shift}
\end{equation}
However the intermediate configuration has zero weight because  $k-1$ is odd, therefore the two determinant ratios in \Eq{eq:shift} are infinity and zero. To eliminate the explicit dependence on the intermediate state with zero weight, we denote
\begin{eqnarray}
	G^{\sigma} (\mathcal{C}_{k}) & = &\left( 
	\begin{array}{cc}
		\tilde{P} & \tilde{Q}\\
		\tilde{R} & \tilde{S} 
	\end{array}
	\right)^{-1}, \, 
	G^{\sigma} (\mathcal{C}_{k} \setminus(\mathbf{i},\tau))   =  M^{-1}, \nonumber \\
	G^{\sigma} (\mathcal{C}_{k}^{\prime})& = &\left(\begin{array}{ccc}
M^{-1} & \multicolumn{1}{r|}{} &  \mathcal{G}^{\sigma}_{\mathbf{i}_{p} \mathbf{i}^{\prime}}(\tau_{p}-\tau^{\prime})  \\ \cline{1-3} 
\mathcal{G}^{\sigma}_{\mathbf{i}^{\prime} \mathbf{i}_{q}}(\tau^{\prime}-\tau_{q})   & \multicolumn{1}{r|}{}& \mathcal{G}^{\sigma}_{\mathbf{i}^{\prime} \mathbf{i}^{\prime} }(0^{+}) -\frac{1}{2}
\end{array}\right) \nonumber \\ &  =&  \left( 
	\begin{array}{cc}
	   M^{-1} & Q\\
		R & S 
	\end{array}
	\right)= \left( 
	\begin{array}{cc}
		\tilde{P}' & \tilde{Q}'\\
		\tilde{R}' & \tilde{S}'
	\end{array}
	\right)^{-1},
\label{eq:Ginvs}
\end{eqnarray}
where $\tilde{P},M,\tilde{P}'$ are $(k-1)\times (k-1)$ matrices, while $\tilde{S},S,\tilde{S}'$ are numbers. 
All Green's function matrices (and their inverses) in \Eq{eq:Ginvs} satisfy the symmetry property \Eq{eq:Gsymmetry}, thus we know that their diagonal elements are all zero, in particular $(\tilde{S},S,\tilde{S}')=(0,0,0)$. We have direct access to $\tilde{P},\tilde{Q},\tilde{R}$  because they are stored for the current configuration, while $Q,R$ can be readily calculated since they are related to the noninteracting Green's function of the to-be-added vertex $(\mathbf{i}^{\prime},\tau^{\prime})$. 
We shall derive a well-defined ratio for \Eq{eq:shift} by eliminating the intermediate state and taking the limit  $(\tilde{S},S,\tilde{S}')\rightarrow(0,0,0)$ in the final step.  
The matrix $M$ is obtained from a removal update\cite{Gull:2011jd}
\begin{equation}
M = \tilde{P} - \tilde{Q}\tilde{S}^{-1}\tilde{R}. 
\end{equation}
The matrices $\tilde{P}',\tilde{Q}',\tilde{R}',\tilde{S}'$ are then obtained from a subsequent insertion update\cite{Gull:2011jd}
\begin{eqnarray}
\tilde{S}' &=& (S-RMQ)^{-1}, \label{eq:S} \\
\tilde{Q}' &=& -MQ\tilde{S}',  \label{eq:Q} \\
\tilde{R}' &=& -\tilde{S}' RM, \label{eq:R} \\
\tilde{P}' &=& M + MQ \tilde{S}'RM. \label{eq:P}
\end{eqnarray}
Equation~(\ref{eq:shift}) can then be calculated as the ratio 
\begin{eqnarray}
\frac{\det \left(G^{\sigma}(\mathcal{C}^{\prime}_{k}) \right)}{\det \left(G^{\sigma}(\mathcal{C}_{k})\right)} &=&\tilde{S}/\tilde{S}' = \tilde{S}(S-RMQ) \nonumber \\ & =& \tilde{S}\left(S-R(\tilde{P} - \tilde{Q}\tilde{S}^{-1}\tilde{R})Q\right)  \nonumber \\ & \rightarrow & (R\tilde{Q} ) (\tilde{R}Q). 
\label{eq:shift_ratio}
\end{eqnarray}
In the last step we have taken the limit 
$\tilde{S}\rightarrow0$. If the shift move is accepted, we calculate the matrices Eqs.~(\ref{eq:Q}-\ref{eq:P}) and store them for the inverse of $G^{\sigma}(\mathcal{C}^{\prime}_{k})$
\begin{align}
\tilde{Q}' 
& =-MQ(S-RMQ)^{-1} 
\rightarrow{\tilde{Q}}/(R\tilde{Q}), \\
\tilde{R}' & = -(S-RMQ)^{-1}RM 
\rightarrow \tilde{R}/(\tilde{R}Q), \\
\tilde{P}' 
& = M+\frac{MQRM}{S-RMQ}\nonumber \\
& \rightarrow\tilde{P}-\tilde{Q}'(R\tilde{P})-(\tilde{P}Q)\tilde{R}'+\tilde{Q}'(R\tilde{P}Q)\tilde{R}'. 
\end{align}
All equations in the above involve only matrix-vector or vector outer-product operations, which have the same computation complexity $\mathcal{O}(k^2)$ compared to the fast update for insertion/removal updates.\cite{Rubtsov:2005iw, Gull:2011jd}

\subsubsection{Measurements}
Because of the vanishing of Monte Carlo weights for odd expansion orders, the configuration space sampled in the CT-INT simulation does not necessarily suffice to measure  all physical observables. In particular, measurements of two-particle correlation functions (such as density-density correlations) need special attention. In this section we present detailed derivation of the Monte Carlo estimators for them in the correlated double-vertex update scheme. Measurement of single-particle quantities such as Green's function follow the standard procedure.\cite{Rubtsov:2005iw, Gull:2011jd}

We write the longitudinal spin correlation as $\hat{S}_{\mathbf{i}}^{z}\hat{S}_{\mathbf{j}}^{z}=\frac{1}{4}\sum_{\sigma\sigma'}\sigma\sigma'\left(\hat{n}_{\mathbf{i}\sigma}-\frac{1}{2}\right)\left(\hat{n}_{\mathbf{j}\sigma^{\prime}}-\frac{1}{2}\right)$, which consists of equal-spin ($\sigma=\sigma'$) and unequal-spin ($\sigma\neq\sigma'$) density-density correlations. The equal-spin correlation for $\mathbf{i}=\mathbf{j}$ is just $1/4$. While for $\mathbf{i}\neq \mathbf{j}$ it could be measured in the usual way by inserting two additional vertices, leading to the estimator\cite{Rubtsov:2005iw, Gull:2011jd}
\begin{align}
\left\langle 
\left(\hat{n}_{\mathbf{i}\sigma}-\frac{1}{2}\right)\left(\hat{n}_{\mathbf{j}\sigma}-\frac{1}{2}\right)
\right\rangle_{\mathcal{C}_{k},\tau} 
=\frac{\det \left( G^{\sigma} (\mathcal{C}_{k+2}) \right) }{\det \left(G^{\sigma} (\mathcal{C}_{k}) \right)}, 
\end{align}
where $G^{\sigma}\left(\mathcal{C}_{k}\right)$ is the Green's function matrix for the current configuration $\mathcal{C}_{k}$, and $\mathcal{C}_{k+2}=\mathcal{C}_{k}\cup(\mathbf{i},\tau)\cup(\mathbf{j},\tau)$ has two more vertices at the same imaginary time $\tau\in[0, \beta)$ which we sample randomly. Because the dimensions of both Green's function matrices are even, their determinants are generally nonzero and the ratio is well-defined. 

On the other hand, it is not so straightforward to measure the unequal-spin correlations. The usual approach would suggest the following estimator
\begin{align*}
 \left\langle 
\left(\hat{n}_{\mathbf{i}\uparrow}-\frac{1}{2}\right)\left(\hat{n}_{\mathbf{j}\downarrow}-\frac{1}{2}\right)
\right\rangle_{\mathcal{C}_{k},\tau} \stackrel{?}{=}& \frac{\det G^{\uparrow}\left(\mathcal{C}_{k}\cup(\mathbf{i},\tau)\right)}{\det G^{\uparrow}\left(\mathcal{C}_{k}\right)} \\ \times& \frac{\det G^{\downarrow}\left(\mathcal{C}_{k}\cup(\mathbf{j},\tau)\right)}{\det G^{\downarrow}\left(\mathcal{C}_{k}\right)}, 
\end{align*}
where the configuration $\mathcal{C}_{k}\cup(\mathbf{i},\tau)$ has one more vertex than the current configuration $\mathcal{C}_{k}$. However, the determinant ratio is zero for even expansion orders because the dimension of $G^{\sigma}\left(\mathcal{C}_{k}\cup(\mathbf{i},\tau)\right)$ is odd; while for odd expansion orders the determinant ratio is \emph{infinite}, but these configurations are never sampled because they have \emph{zero} weight. The correct estimator is the latter \emph{zero times infinity} contribution. 

To resolve the problem of measuring the unequal spin correlations, we use the ``shift'' rather than the ``insertion'' measurement.\cite{Kozik:2013ji} The idea is to view $\left(\hat{n}_{\mathbf{i}\uparrow}-\frac{1}{2}\right)\left(\hat{n}_{\mathbf{j}\downarrow}-\frac{1}{2}\right)$ as an existing interaction vertex with a shifted site from $\mathbf{i}$ to $\mathbf{j}$ for the spin down component. To this end, we expand the unequal-spin density correlation observable similar to \Eq{eq:ctint} and use the translational symmetry in space and imaginary time,  
\begin{eqnarray}
&& \left\langle 
\left(\hat{n}_{\mathbf{i}\uparrow}-\frac{1}{2}\right)\left(\hat{n}_{\mathbf{j}\downarrow}-\frac{1}{2}\right)
\right\rangle \label{eq:shift_z_correlation} \\
&=&  \frac{Z_{0}}{Z} \sum_{k=1}^{\infty}\left(-U\right)^{k-1}   \sum_{\mathcal{C}_{k-1}} \times \nonumber   \\ &&  \det G^{\uparrow}\left(\mathcal{C}_{k-1}\cup(\mathbf{i},\tau) \right)  \det G^{\downarrow}\left(\mathcal{C}_{k-1}\cup(\mathbf{j},\tau)\right) \nonumber \\
&=&\frac{-1}{U\beta N} \frac{Z_{0}}{Z} \sum_{k=1}^{\infty}  \left(-U\right)^{k} \sum_{\mathcal{C}_{k-1}} \sum_{\mathbf{i}_{k}} \int_{0}^{\beta} d\tau_{k} \times \nonumber \\
&& \det G^{\uparrow}\left(\mathcal{C}_{k-1}\cup(\mathbf{i}_{k},\tau_{k})\right)  \det G^{\downarrow}\left(\mathcal{C}_{k-1}\cup(\mathbf{i}_{k}+\mathbf{j}-\mathbf{i},\tau_{k})\right)\nonumber. 
\end{eqnarray}
The contribution to the above sum is nonzero only for \emph{even} $k$. Considering $\mathcal{C}_{k}\equiv\mathcal{C}_{k-1}\cup(\mathbf{i}_{k}, \tau_{k})$ as a Monte Carlo configuration sampled with a non-vanishing Monte Carlo weight, the configuration $\mathcal{C}_{k-1}\cup(\mathbf{i}_{k}+\mathbf{j}-\mathbf{i},\tau_{k})$ can be reached by shifting a spin down vertex in space. 
Combining the integration over $\tau_{k}$ with the other $k-1$ time-ordered integrations over the imaginary times, we arrive at the following estimator for the unequal-spin density correlation\cite{Kozik:2013ji}
\begin{align}
\left\langle 
\left(\hat{n}_{\mathbf{i}\uparrow}-\frac{1}{2}\right)\left(\hat{n}_{\mathbf{j}\downarrow}-\frac{1}{2}\right)
\right\rangle_{\mathcal{C}_{k},\tau_{p}} =\frac{-k}{U\beta N}\frac{\det G^{\downarrow}\left(\mathcal{C}^{\prime}_{k}\right)}{\det G^{\downarrow}\left(\mathcal{C}_{k}\right)},
\label{eq:shiftmeasure}
\end{align}
where the configuration $\mathcal{C}^{\prime}_{k}=\mathcal{C}_{k}\setminus(\mathbf{i}_{p},\tau_{p})\cup (\mathbf{i}_{p}+\mathbf{j}-\mathbf{i},\tau_{p}) $ is obtained from $\mathcal{C}_{k}$ by randomly selecting an existing vertex $(\mathbf{i}_{p}, \tau_{p})$ and shifting the site index $\mathbf{i}_{p}$ (of spin down only) to $\mathbf{i}_{p}+\mathbf{j}-\mathbf{i}$. The determinant ratio is calculated by the fast-update formula \Eq{eq:shift_ratio}. The estimator (\ref{eq:shiftmeasure}) is zero when there is no vertex. It  also automatically covers the case $\mathbf{i}=\mathbf{j}$, where the determinant ratio is one and the equation (\ref{eq:shiftmeasure}) reduces to the estimator for the interaction energy.\cite{Rubtsov:2005iw} 
A similar but inequivalent estimator can be obtained by shifting the spin up vertex while fixing the spin down vertex. 
The transverse spin correlation $\braket{\hat{S}_{\mathbf{i}}^{x}\hat{S}_{\mathbf{j}}^{x}}$ can be measured in a similar sprit and will be discussed in Appendix~\ref{appendix:inplane}.

\subsection{Lattice continuous-time auxiliary field algorithm (LCT-AUX)} 
The LCT-AUX approach\cite{1999PhRvL..82.4155R, Iazzi:2014vv} treats the expansion in  \Eq{eq:expansion} as a weighted sum of partition functions of imaginary-time dependent free fermions. To achieve this goal, we first perform an auxiliary field decomposition for the local interaction term \Eq{eq:v},\cite{Wang:2015vw}
\begin{align}
-\hat{v}_{\mathbf{i}}
=\frac{\Gamma}{2} \sum_{s =\pm1} \exp \left[s \lambda \left(\hat{c}_{\mathbf{i}\uparrow}^{\dagger} \hat{c}_{\mathbf{i}\downarrow}+ \hat{c}_{\mathbf{i}\downarrow}^{\dagger} \hat{c}_{\mathbf{i}\uparrow} \right)\right] \label{eq:auxfield+} 
\end{align}
where $\lambda = \mbox{acosh}(1+\frac{U}{2\Gamma})$. This unconventional decomposition, which introduces an auxiliary field that couples to the local spin flip rather than the density or magnetization, is necessary to avoid the sign problem.\cite{Wang:2015vw} Since the auxiliary field couples the two spin species, we introduce a combined spin-orbital index $i=(\mathbf{i}\sigma)$ and write the free Hamiltonian \Eq{eq:H0} as $\hat{H}_{0}=\sum_{ij} \hat{c}_{i}^{\dagger}K_{ij}\hat{c}_{j}$, where $K$ is a $2N\times 2N$ matrix. 
Substituting this and \Eq{eq:auxfield+} into \Eq{eq:expansion} and tracing out the free fermions, one obtains 
\begin{eqnarray}
Z &=& \sum_{k=0}^{\infty} \left(\frac{\Gamma}{2}\right)^{k}\sum_{\mathbf{i}_{1}, \ldots, \mathbf{i}_{k}} \sum_{s_{1},\ldots,s_{k}} \int_{0}^{\beta}d \tau_{1} \cdots \int^{\beta}_{\tau_{k-1}} d\tau_{k}  \nonumber\\
&\times&\det\left[I +e^{-(\beta-\tau_{k})K}{X^{s_{k}}_{\mathbf{i}_{k}}}\cdots {{X^{s_{1}}_{\mathbf{i}_{1}}} } e^{-\tau_{1} K}\right] \nonumber \\
& = & \sum_{k=0}^{\infty} \left(\frac{\Gamma}{2}\right)^{k}\sum_{\mathcal{C}_{k}} \det\left[ I + M(\mathcal{C}_{k}) \right]. \label{eq:weight} 
	\end{eqnarray}
Compared to the CT-INT approach \Eq{eq:ctint}, here the Monte Carlo configuration $\mathcal{C}_{k}=\left\{\left(\mathbf{i}_1,\tau_1,s_1\right),\ldots,\left(\mathbf{i}_k,\tau_k,s_k\right)\right\}$ contains an additional Ising auxiliary field variable $s_\ell$ at each vertex. Moreover, the Monte Carlo weight is given by a single determinant with a fixed matrix size $2N\times 2N$ instead of two matrices of size $k\times k$ for the two spin components. The vertex matrix $X^{s}_{\mathbf{i}}$ has a form following directly from~\Eq{eq:auxfield+}, 
\begin{eqnarray}
X^{s}_{\mathbf{i}} &= &   
\left(
\begin{array}{cccc}
   I &  & &  \\
   &\cosh(s\lambda) & \sinh(s\lambda)  \\
   &\sinh(s\lambda) & \cosh(s\lambda)  \\
   & &  & I   
\end{array} 
\right) \label{eq:Xm1}  \\ 
&=&  \left(
\begin{array}{cccc}
   I &  & & \\
   &1+\frac{U}{2\Gamma}& s\sqrt{\frac{U}{\Gamma}(1+\frac{U}{4\Gamma})}   \\
   &s\sqrt{\frac{U}{\Gamma}(1+\frac{U}{4\Gamma})}& 1+\frac{U}{2\Gamma}  \\
   & &  &I 
\end{array} 
\right). \nonumber  
\end{eqnarray}
It differs from the identity matrix only in the $2\times 2$ block that involves $\mathbf{i}\uparrow$ and $\mathbf{i}\downarrow$. 

\subsubsection{Absence of the sign problem}
The absence of sign problem in~\Eq{eq:weight} is due to a remarkable Lie group property of the evolution matrix $M(\mathcal{C}_{k})= e^{-(\beta-\tau_{k})K}{X^{s_{k}}_{\mathbf{i}_{k}}}\cdots {{X^{s_{1}}_{\mathbf{i}_{1}}} } e^{-\tau_{1} K}$ in the Monte Carlo weight.\cite{Wang:2015vw} To reveal it we define a diagonal matrix $D$ whose nonzero elements read $D_{\mathbf{i}\sigma,\mathbf{i}\sigma}=\eta_{\mathbf{i}}\sigma$. These diagonal elements contain $N$ of ones and $N$ of minus ones, thus provide an indefinite metric.  
One can readily see that for the choice $\Gamma\in [-U/4, 0)\cup(0, \infty)$ the evolution matrix $M$ is real valued and $M^{T} D M = D$, since each factor of $M$ satisfies the same condition. $M$ thus belongs to the split orthogonal group which contains four disconnected components. Remarkably, the determinant $\det(I + M)$ has a definite sign (or vanishes) for each component and there is no sign problem for any shift $\Gamma\in [-U/4, 0)\cup(0, \infty)$.\cite{Wang:2015vw} 
In particular, for the special choice of $\Gamma = -U/4$, the vertex matrix \Eq{eq:Xm1} becomes diagonal and the weight in \Eq{eq:weight} is proportional to \Eq{eq:ctint}. Hence all odd expansion orders have vanishing weight and the above formalism reduces to the LCT-INT approach.\cite{Iazzi:2014vv,Wang:2015tf} 
The choice of $\Gamma$ will affect the efficiency of the simulation because the average expansion order $\braket{k} = -\beta \sum_{\mathbf{i}}\braket{\hat{v}_{\mathbf{i}}} $ increases linearly with $\Gamma$. In the following simulation we choose $\Gamma = -U/4 + 0.05$ to leave finite Monte Carlo weights for odd expansion orders, such that the complications in the Monte Carlo updates and measurements as in the CT-INT method, are avoided. 

 
\subsubsection{Monte Carlo updates} 
The Monte Carlo simulation consists of randomly insertion or removal of vertex matrices into \Eq{eq:weight}. We refer the reader to Refs.~\onlinecite{Iazzi:2014vv,Wang:2015tf} for the general procedure of efficient and stable QMC simulation. In the following we highlight the key steps. The central quantity of the LCT-AUX simulation is the equal-time Green's function calculated for a given configuration $\braket{\hat{c}_{i}\hat{c}_{j}^\dagger}_{\mathcal{C}_{k}, \tau}={G_{ij}}$. To express it in terms of the evolution matrices, we split the matrix product at the imaginary time $\tau$ and write $M(\mathcal{C}_{k})= L R$ such that $R$ denote the matrix product from $0$ to $\tau$ and $L$ from $\tau$ to $\beta$ respectively. The Green's function is $G = (I+R L)^{-1}$.\cite{Loh:1992}

To calculate the acceptance rate of a vertex insertion $(\mathbf{i}, \tau, s)$, one needs to calculate the determinant ratio
\begin{equation}
\frac{\det( I +L X^{s}_{\mathbf{i}}R)}{\det(I+LR)}=\det\left[I+(X^{s}_{\mathbf{i}}-I)(I-G)\right]. 
\label{eq:lct-aux-ratio}
\end{equation}
Since $X^{s}_{\mathbf{i}}-I$ are nonzero only in the entries involving $\mathbf{i}\uparrow$ and $\mathbf{i}\downarrow$, the determinant ratio calculation only involves a $2\times2$ block of the matrix $G$. If the insertion is accepted, the matrix $G$ is updated according to the Woodbury matrix identity\cite{woodbury1950inverting}
\begin{eqnarray}
G' &=& (I+X^{s}_{\mathbf{i}} R L)^{-1} \label{eq:lct-aux-updateG} \\ 
&=& L^{-1}\frac{1}{\left(I+ LR \right) + L\left(X^{s}_{\mathbf{i}}-I\right)R}L  \nonumber \\
&=& G-G\mathcal{P} \left[ \frac{1}{\mathcal{P}^{T}\left( (X^{s}_{\mathbf{i}}-I)^{-1}+ (I-G) \right ) \mathcal{P}} \right]  \mathcal{P}^{T} (I-G), \nonumber
\end{eqnarray}
which again only depends on $G$ but not on the detailed information of $L$ and $R$. 
In particular, the projector $\mathcal{P}$ is a $2N\times 2$ matrix that projects to the nonzero block of the matrix $X^{s}_{\mathbf{i}}-I$. The update can thus be evaluated with $\mathcal{O}(N^{2})$ operations. To remove a vertex $(\mathbf{i}, \tau, s)$, we use the same formulae Eqs.~(\ref{eq:lct-aux-ratio}-\ref{eq:lct-aux-updateG}) except now the matrix $(X^{s}_{\mathbf{i}})^{-1}=X^{-s}_{\mathbf{i}}$ is inserted at time $\tau$ to cancel the existing vertex matrix.   


\subsubsection{Measurements} 
Measurements in LCT-AUX can be performed based on Wick contractions of equal-time Green's functions $G$.~\footnote{They are calculated at a random imaginary time during the sweep.}  Since the auxiliary field couples the two spin components, one needs to take into account additional Wick contractions between different spins components. For example, the estimator for density-density correlation is 
\begin{equation} 
 \braket{\hat{n}_{i} \hat{n}_{j} }_{\mathcal{C}_{k},
 \tau} =  ( G_{i i}-1)( G_{j j}-1) - G_{i j}(  G_{j i} -\delta_{ji}),   
\end{equation} 
where we are still using the combined indices $i=(\mathbf{i}\sigma)$ and $j=(\mathbf{j}\sigma^{\prime})$.
While a general two-body correlations follow
\begin{align}
\braket{\hat{c}_{i}\hat{c}_{j}^\dagger\hat{c}_{k}\hat{c}_{l}^\dagger}_{\mathcal{C}_{k},\tau}={G_{ij}G_{kl}-G_{il}(G_{kj}-\delta_{kj})}. 
\end{align}
For the calculation of the Binder ratio we need to calculate 
$M_{4}$ in \Eq{eq:M4}, which involves four density correlations 
\begin{eqnarray}
 &&  \braket{
  \hat{n}_{i} \hat{n}_{j} \hat{n}_{k} \hat{n}_{l} }_{\mathcal{C}_{k},\tau} \\
 & = &\det \left(\begin{array}{cccc}
    G_{i i} - 1 & G_{i j} & G_{i k} & G_{i l}\\
    G_{j i}  - \delta_{j i}  & G_{j j} - 1 & G_{j k} & G_{j l}\\
    G_{k i} - \delta_{k i} & G_{k j} - \delta_{k j} & G_{k k} - 1 & G_{k
    l}\\
    G_{l i} - \delta_{l i} &  G_{l j} - \delta_{l j} & 
    G_{l k} - \delta_{l k} & G_{l l} - 1
  \end{array}\right). \nonumber
\end{eqnarray}
Compared to CT-INT, measuring these equal-time correlation functions is much easier in LCT-AUX because there is no subtle \emph{zero times infinity} problem caused by the vanishing of Monte Carlo weights of odd expansion orders.

\section{Results \label{sec:results}}
We first present benchmark results in Sec.~\ref{sec:benchmark} to demonstrate the correctness of the implementations, then present results on spin correlations on a one-dimensional chain in Sec.~\ref{sec:1dchain}, and the thermal phase transition on the square lattice in Sec.~\ref{sec:square}. Finally, we report results on the quantum phase transition on the honeycomb lattice in Sec.~\ref{sec:honeycomb}.

\subsection{Benchmarks \label{sec:benchmark}}
\begin{figure}[t]
\includegraphics[width=8cm]{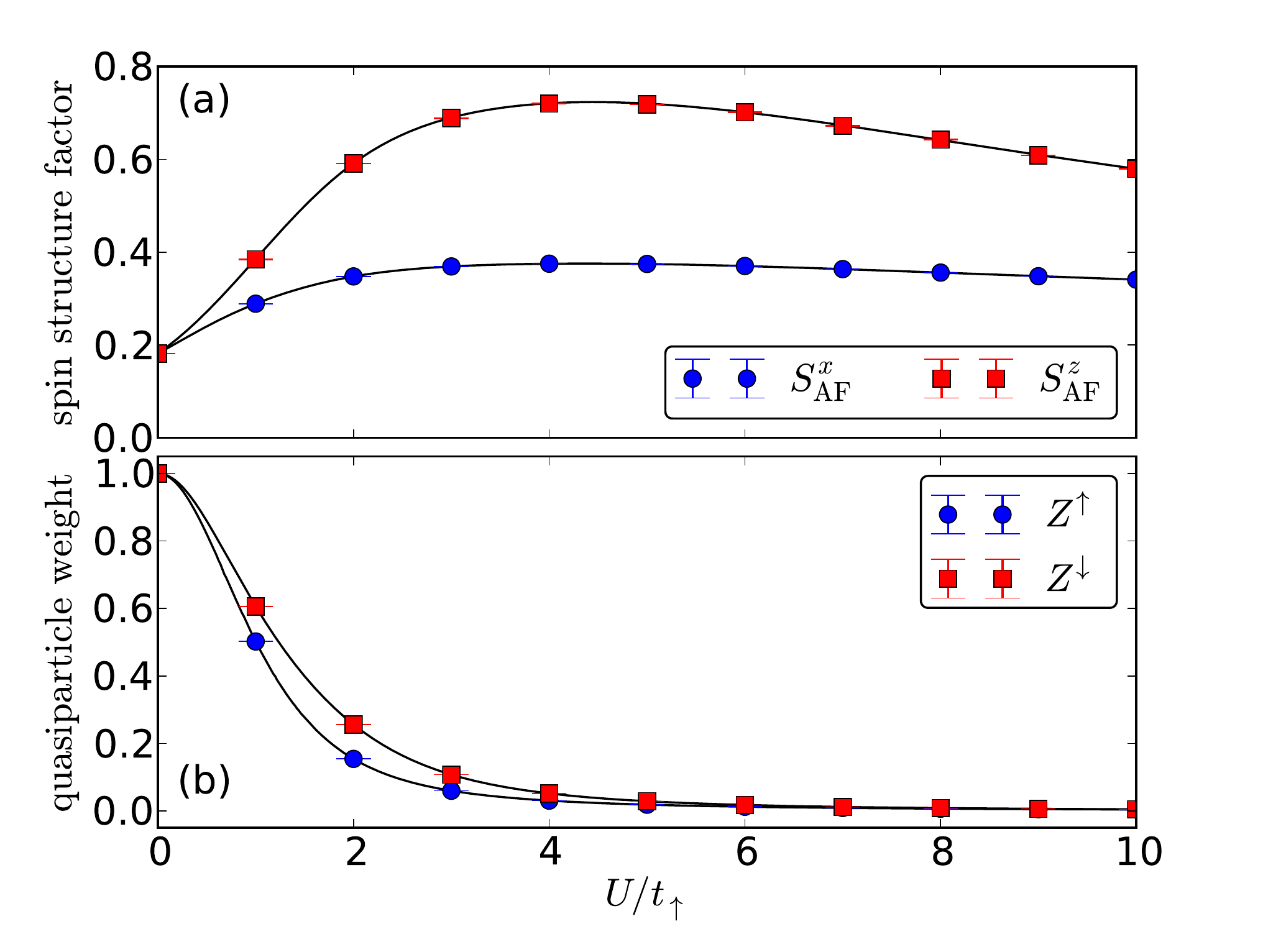}
\caption{Benchmark of CT-QMC (markers) with exact diagonalization (black lines) for (a) spin structure factors and (b) quasiparticle weights. Calculations are performed on a four-site periodic Hubbard chain and at temperature $\beta t_\uparrow=10$ and mass imbalance $t_\downarrow/t_\uparrow=0.15$. 
\label{fig:bench1}}
\end{figure}

Figure~\ref{fig:bench1}(a) compares the QMC results on the spin structure factor with the exact diagonalization for a four-site Hubbard chain with the periodic boundary condition. The results obtained by the two CT-QMC methods fully agree with the exact results. Furthermore, as expected, the longitudinal antiferromagnetic structure factor is larger than the transverse one for general asymmetric cases.

\begin{figure}[t]
\includegraphics[width=8cm]{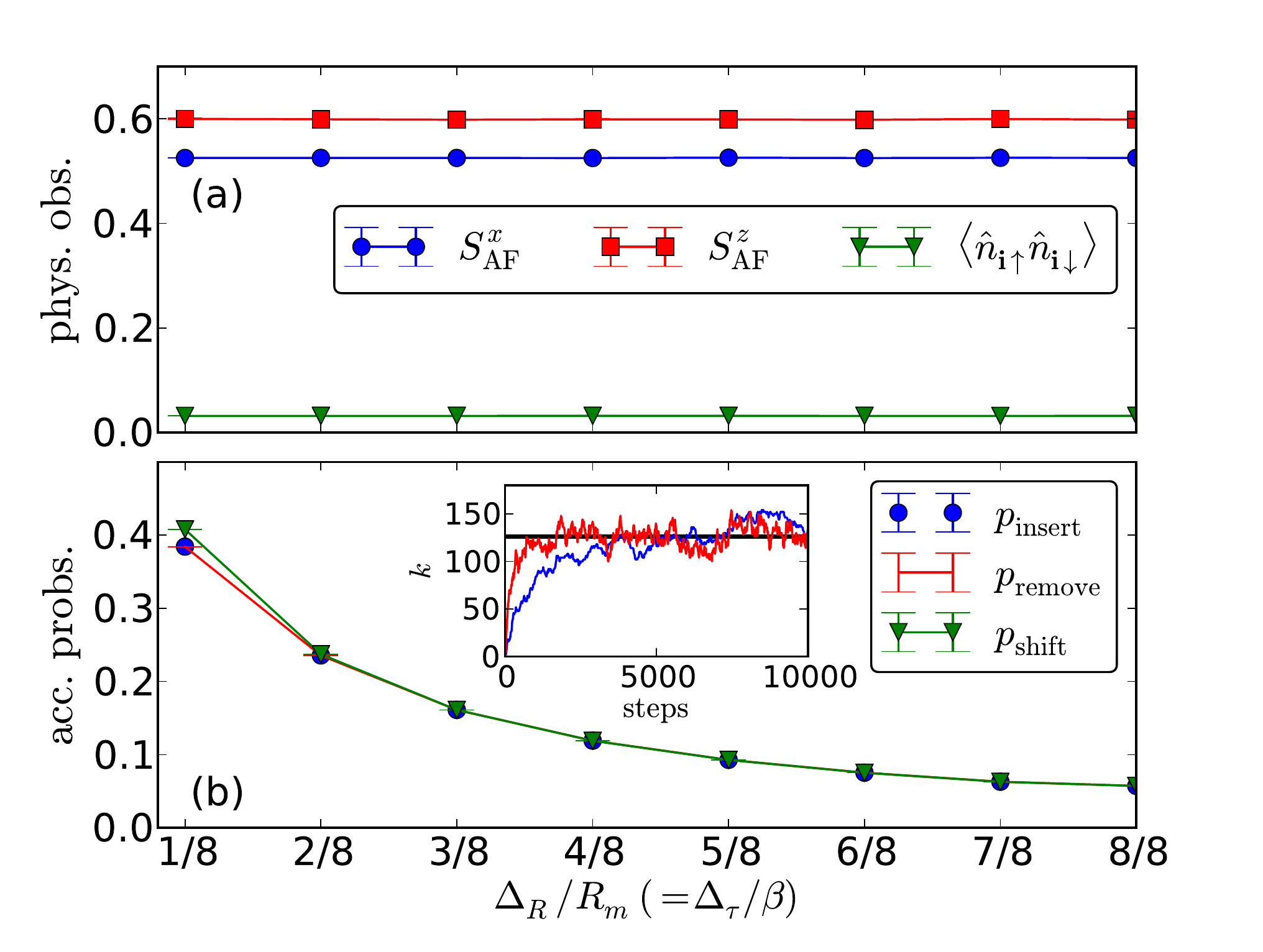}
\caption{Effects of correlated double-vertex update in CT-INT. 
Dependence of (a) the physical observables and (b) the acceptance probabilities on the cutoffs $\Delta_R$ and $\Delta_\tau$. Calculations are performed on the chain lattice: $L=16$, $\beta t_\uparrow=6$, $U/t_\uparrow=6$, and $t_\downarrow/t_\uparrow=0.5$. We have set $\Delta_R/R_m=\Delta_\tau/\beta$. As shown in (a), the spin structure factors and double occupancy do not depend on these ratios; while shown in (b), the acceptance probabilities are greatly enhanced by the  correlations between the inserted/removed vertices. The initial thermalization processes for $\Delta_R/R_m=\Delta_\tau/\beta=1/8$ (red) and $\Delta_R/R_m=\Delta_\tau/\beta=1$ (blue) are shown in the inset, which indicate correlated double-vertex update indeed accelerates the Monte Carlo thermalization significantly. 
\label{fig:bench2}}
\end{figure}

We then calculate the quasiparticle weight at the Fermi surface for each spin, which measures the enhancement of the quasiparticle mass due to correlation effects. It is approximately obtained by an analytical continuation on the imaginary-frequency axis as
\begin{eqnarray}
Z^\sigma&\approx&\left[1-\mathrm{Im}\Sigma^\sigma(\mathbf{k}_F,i\omega_0)/{\omega_0}\right]^{-1},
\end{eqnarray}
where $\Sigma^{\sigma}$ is the self-energy, $\mathbf{k}_F$ is the Fermi wavevector (on a chain $k_F=\pi/2$) and $\omega_0=\pi T$ is the first fermionic Matsubara frequency. 
From Fig.~\ref{fig:bench1}(b), the quasiparticle weights are indeed suppressed at larger interaction strengths, while the spin-down component with smaller Fermi velocity has larger weight. A similar phenomenon has been observed in dynamical mean-field theory calculations on the Bethe lattice.\cite{Winograd:2011en,PhysRevA.85.013606} Here the Fourier transform of the interacting Green's function (needed to calculate the self energy) is obtained using the CT-INT estimator derived in \Ref{Rubtsov:2005iw}, which also applies to rank-2 updates.

The LCT-AUX method scales as $\mathcal{O}(\beta U N^{3})$ compared to the $\mathcal{O}(\beta^{3}U^{3}N^{3})$ scaling of the CT-INT methods. Thus LCT-AUX is  asymptotically better for reaching low temperature (or the ground state) and dealing with strong interactions. Whenever both methods are applicable, we find that they give the same results within statistical errors. 

We next present further technical results of the two methods. 
Figure~\ref{fig:bench2} shows the effects of correlated double-vertex update parameters $\Delta_R$ and $\Delta_\tau$ in CT-INT simulations. To this end, we simulate an $N=16$ chain lattice with periodic boundary conditions, setting the ratios $\Delta_R/R_m$ and $\Delta_\tau/\beta$ equal, and varying them from $1/8$ to $1$, where $R_m = 8$ is the maximal distance on the chain lattice of the length $N=16$. As expected, all physical results are independent of the parameters $\Delta_{R}$ and $\Delta_{\tau}$. However the correlated double-vertex update increases the acceptance probabilities by almost an order of magnitude. The effect is more dramatic in larger systems at lower temperatures where the simple double-vertex updates have even lower acceptance rate. Note that the acceptance probabilities for insertion and removal are the same in equilibrium. The inset of Fig.~\ref{fig:bench2}(b) shows the expansion order in the equilibration phase of the simulation. One clearly observes that the correlated double-vertex update increases the efficiency of the CT-INT simulation.

\begin{figure}[t]
\includegraphics[width=8cm]{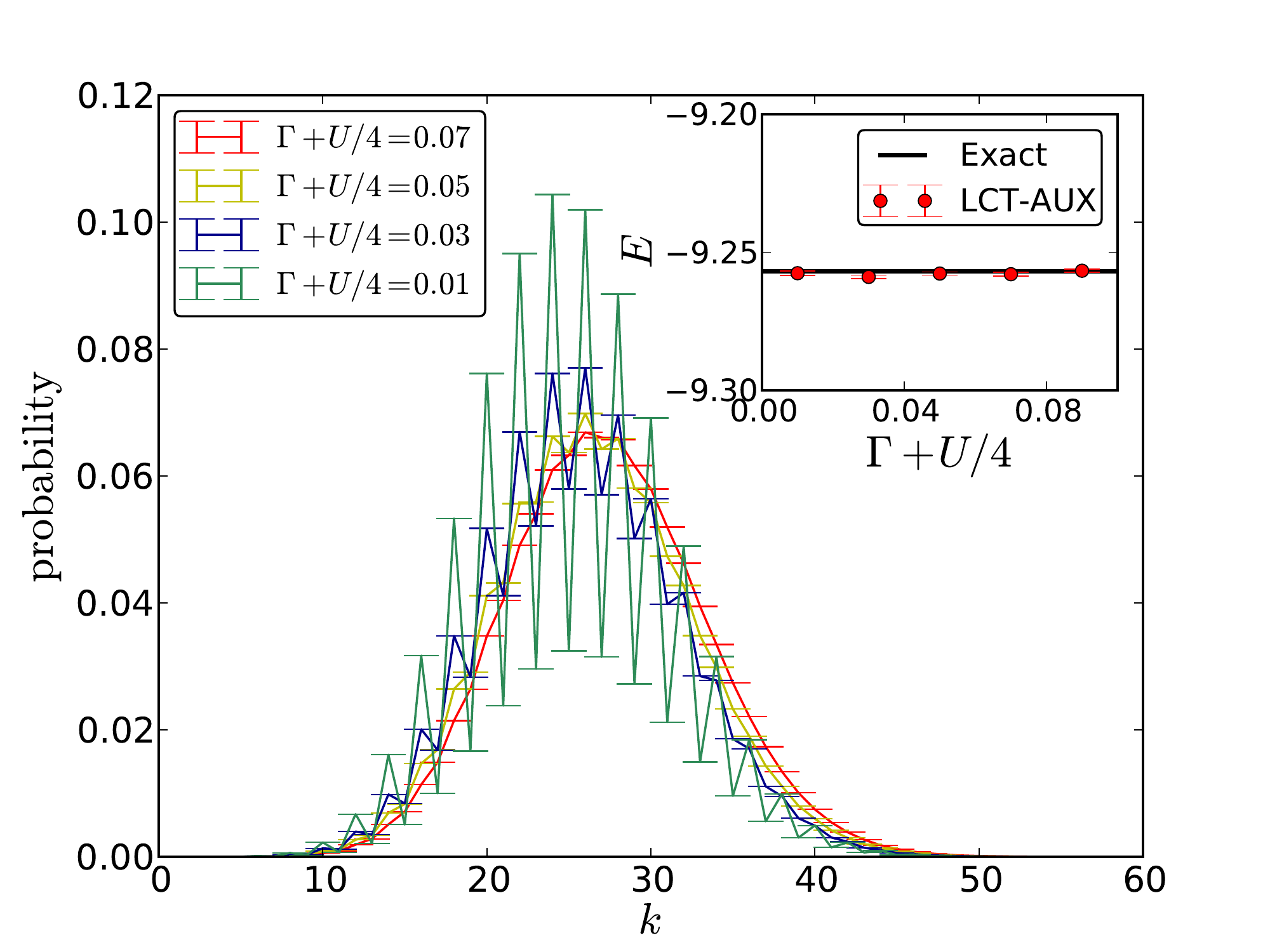}
\caption{Effects of the constant shift $\Gamma$ in LCT-AUX. 
The histograms of the expansion order with various choices of $\Gamma$. Results are obtained on a four-site chain with $t_{\downarrow}/t_{\uparrow}=0.5$, $U/t_{\uparrow}=4$ and $\beta t_{\uparrow}=8$. Inset demonstrates that physical observables such as the total energy (of the Hamiltonian \Eq{eq:Ham}) is independent of $\Gamma$. The black solid line shows the exact value of total energy. 
\label{fig:histograms}}
\end{figure}

Figure~\ref{fig:histograms} shows the histogram of the expansion order in LCT-AUX for various choices of $\Gamma$. One clearly sees the weights of odd expansion orders get suppressed for $\Gamma$ close to $-U/4$. Increasing $\Gamma$  enhances their weights but also increases the average expansion order. Inset shows that physical observables, such as the total energy does not depend on the value of the shift. We choose $\Gamma = -U/4+0.05$ throughout this paper for LCT-AUX calculations.  

\begin{figure}[t]
\includegraphics[width=8cm]{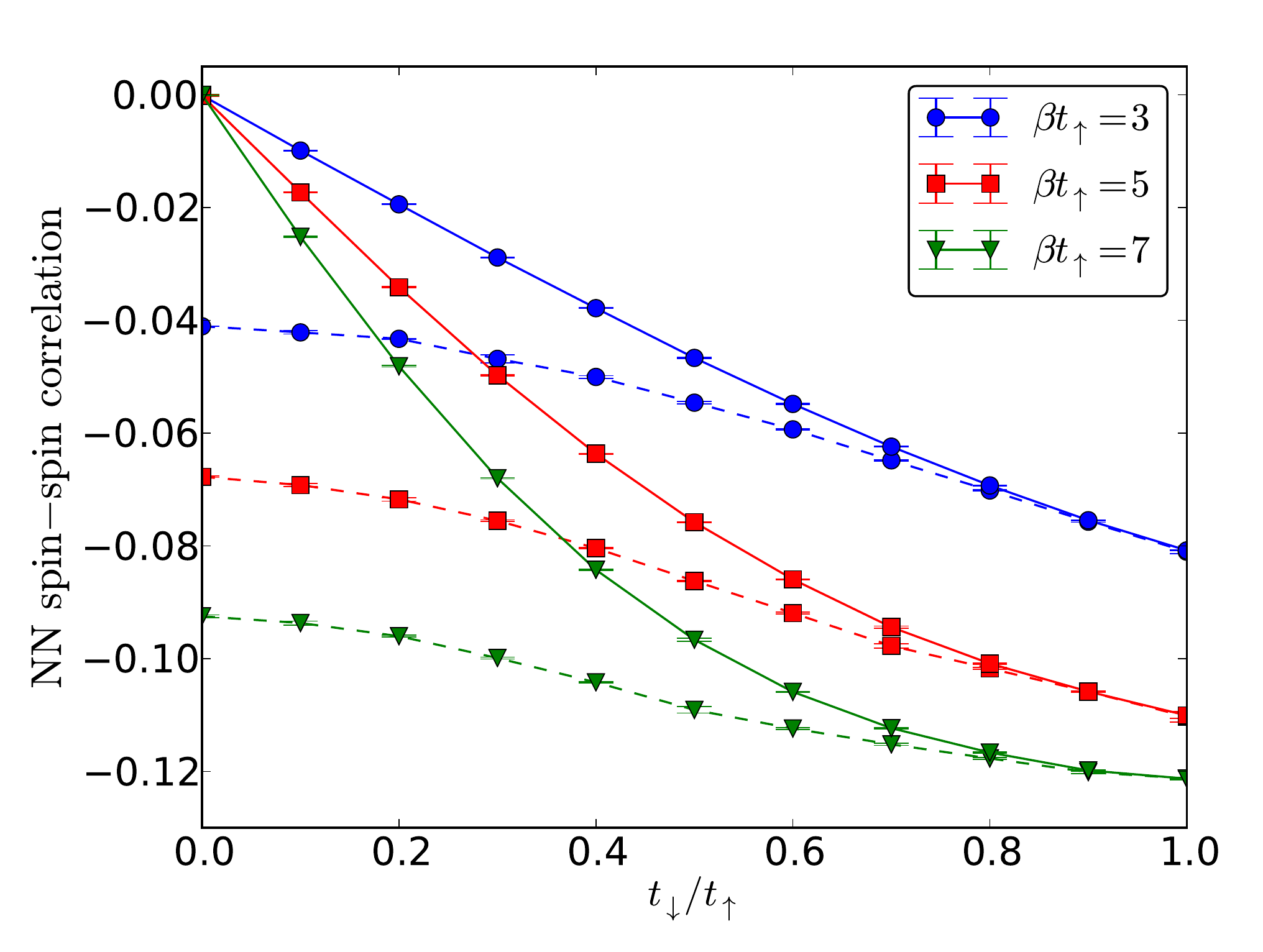}
\caption{Nearest-neighbor spin-spin correlations on a $L=32$ periodic chain at $U=8t_\uparrow$. Solid lines represent $\braket{\hat{S}_\mathbf{i}^{x}\hat{S}_{\mathbf{i}+\hat{x}}^{x}}$ while dashed ones $\braket{\hat{S}_\mathbf{i}^{z}\hat{S}_{\mathbf{i}+\hat{x}}^{z}}$. When $t_\uparrow=t_\downarrow$, the spin correlations in both directions are equal. When $t_\downarrow=0$, the longitudinal spin correlation saturates to a finite value while the transverse spin correlation vanishes. This agrees with the large-$U$ analysis based on the XXZ model \Eq{eq:xxz}. 
\label{fig:chain1}} 
\end{figure}

\subsection{Spin correlations of a one-dimensional chain~\label{sec:1dchain}}

Figure~\ref{fig:chain1} shows the nearest-neighbor spin-spin correlations calculated in a periodic chain of length $L=32$ and $U=8t_{\uparrow}$ at various temperatures. Both the longitudinal and the transverse spin correlations are negative, indicating antiferromagnetic correlations between nearest neighbors. However,  they are equal only at $t_\uparrow=t_\downarrow$. The transverse spin correlation decreases and vanishes as $t_\downarrow\rightarrow0$, which again agrees with the large-$U$ understanding, because $J_{xy}$ vanishes. The longitudinal spin correlation, on the other hand, saturates to a value in the limit of $t_\downarrow\rightarrow0$ since $J_z\neq 0$ in this limit.  Overall, all spin correlations are enhanced at lower temperatures due to the suppression of thermal fluctuations.

These predictions can be verified in a recently experimental realization of the one-dimensional mass-imbalanced Hubbard model.\cite{Jotzu:2015tq}

\subsection{Thermal phase transition on square lattice \label{sec:square}}

In Fig.~\ref{fig:phasediag}(a) we sketch the finite temperature phase diagram for model~(\ref{eq:Ham}) (for a fixed $U$) on the square lattice. A crucial consequence of the reduced symmetry of model~(\ref{eq:Ham}) is that, the system can develop the long range antiferromagnetic Ising order at finite temperatures because a discrete $Z_{2}$ symmetry is broken. The transition temperature drops to zero (red dot) in the symmetric hopping case $t_{\uparrow}=t_{\downarrow}$, as required by the Mermin-Wagner theorem for the spontaneous breaking of the continuous $SU(2)$ symmetry.\cite{PhysRevLett.17.1133} 

\begin{figure}[t]
\includegraphics[width=8cm]{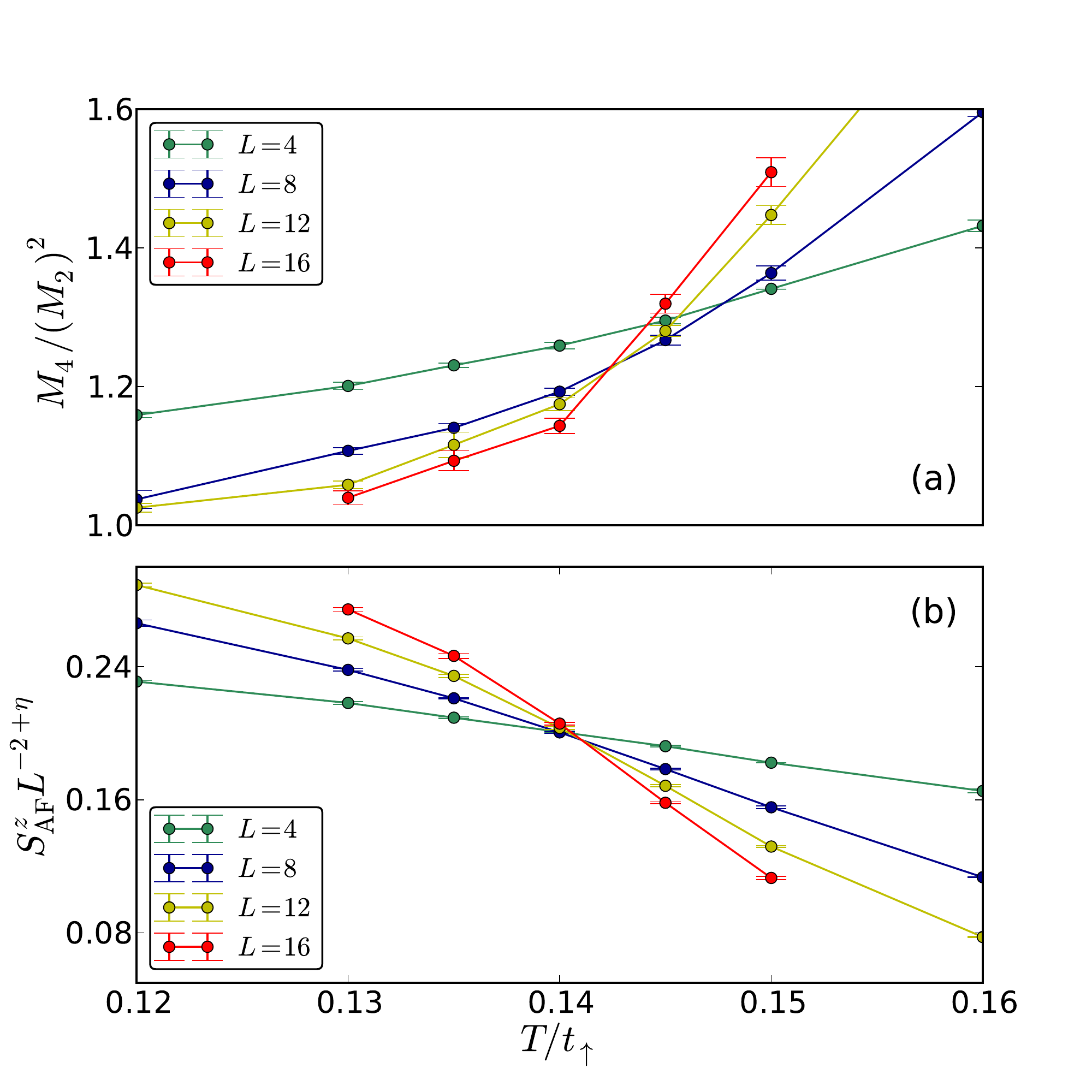}
\caption{(a) Binder ratio (b) scaled staggered spin structure factor versus temperature on the square lattice with $t_{\downarrow}/t_{\uparrow}=0.15$ and $U/t_{\uparrow}=4$. The crossing point indicates the critical temperature to an antiferromagnetic Ising state. 
\label{fig:Ising}}
\end{figure}

The exact value of the transition temperature also depends on the interaction strength $U/t_{\uparrow}$. In the Falicov-Kimball limit, it attains a maximal value $T/t_{\uparrow}\approx 0.15$ at $U/t_{\uparrow} = 4$ (purple dot).\cite{Maska:2006id} As a representative for general mass-imbalanced cases, we consider $t_{\downarrow}/t_{\uparrow}=0.15$ corresponding to the mass ratio of $^{6}$Li and $^{40}$K atoms and $U/t_{\uparrow} = 4$. Figure~\ref{fig:Ising}(a) shows the Binder ratio as a function of temperature which has a crossing at the critical temperature. The value of the Binder ratio at the crossing point approaches to the universal value $1.1679$ of the 2D Ising model on a torus with isotropic couplings.\cite{Salas:1999qh}  Figure~\ref{fig:Ising}(b) shows the scaled spin structure factor according to the 2D Ising critical exponent $\eta = 0.25$, which also gives consistent transition temperature $T_{c}/t_{\uparrow}\approx 0.142$. These numerical data are consistent with the expected 2D Ising universality class of the thermal phase transition of model~(\ref{eq:Ham}) on the square lattice. 



%

\subsection{Quantum phase transition on honeycomb lattice \label{sec:honeycomb}}
As a more challenging application, we finally study the quantum phase transition of the mass-imbalanced Hubbard model~(\ref{eq:Ham}) on the honeycomb lattice. As illustrated in Fig.~\ref{fig:phasediag}(b) the single-particle band structure features two sets of Dirac cones in the Brillouin zone. Each set consists of two Dirac cones with unequal Fermi velocities due to the hopping asymmetry. In the $SU(2)$ case $t_\uparrow=t_\downarrow$,\cite{Sorella:1992wd, Meng:2010gc} it is known that there is a continuous quantum phase transition from the Dirac semimetal to an antiferromagnetic Heisenberg insulator at $U \approx 3.8t_\uparrow$ (red dot), which is well described by the Gross-Neveu model.\cite{Herbut:2006jaa, Sorella:2012hib, Assaad:2013kg} On the other hand, at the Falicov-Kimball limit where $t_{\downarrow}=0$, the system shows antiferromagnetic Ising order at arbitrarily small repulsion (purple dot).\cite{Kennedy:1986wf} The exact phase boundary for general asymmetric hoppings remains open. One nevertheless anticipates a finite critical interaction strength, since the density of states at the Fermi level is still zero. Furthermore, because of the reduced symmetry compared to the $SU(2)$ Hubbard model, we anticipate the critical behavior of the transition, from an spin-splitted Dirac semimetal to the antiferromagnetic Ising insulator, is different from the Gross-Neveu model with the $SU(2)$ symmetry.


To directly address these questions at zero temperature, we employ a projector version of the LCT-AUX algorithm. We sample the configurations not from the partition function but from the wavefunction overlap $\braket{\Psi_{T} |e^{-\beta\hat{H}} |  \Psi_{T}}$, where $\ket{\Psi_{T}} = \prod_{n=1}^{N} \left(\sum_{i=1}^{2N} P_{in} \hat{c}^{\dagger}_{i}\right) \Ket{\mathrm{vac}}$ is a trial wavefunction which we choose to be the ground state of the free Hamiltonian~(\ref{eq:H0}), i.e. $P$ contains the occupied eigenvectors of the single particle hopping matrix $K$. The Monte Carlo weight in \Eq{eq:weight} thus becomes $w(\mathcal{C}_{k}) = \det\left(P^{\dagger} e^{-(\beta-\tau_{k})K}{X^{s_{k}}_{\mathbf{i}_{k}}}\cdots {{X^{s_{1}}_{\mathbf{i}_{1}}} } e^{-\tau_{1} K}  P \right)$. Physical observables are measured at the center of the projection $\tau=\beta/2$. Since the acceptance rate and updates follow the same equations as described in Eqs.~(\ref{eq:lct-aux-ratio}-\ref{eq:lct-aux-updateG}), there is no sign problem in the zero temperature simulation either. Below we report results for hopping asymmetry $t_{\downarrow}/t_{\uparrow}=0.15$, system size $N=2L^{2}$ with $L=6,9,12,15$, and projection time $\beta t_{\uparrow}=40$. 



\begin{figure}[t]
\includegraphics[width=8cm]{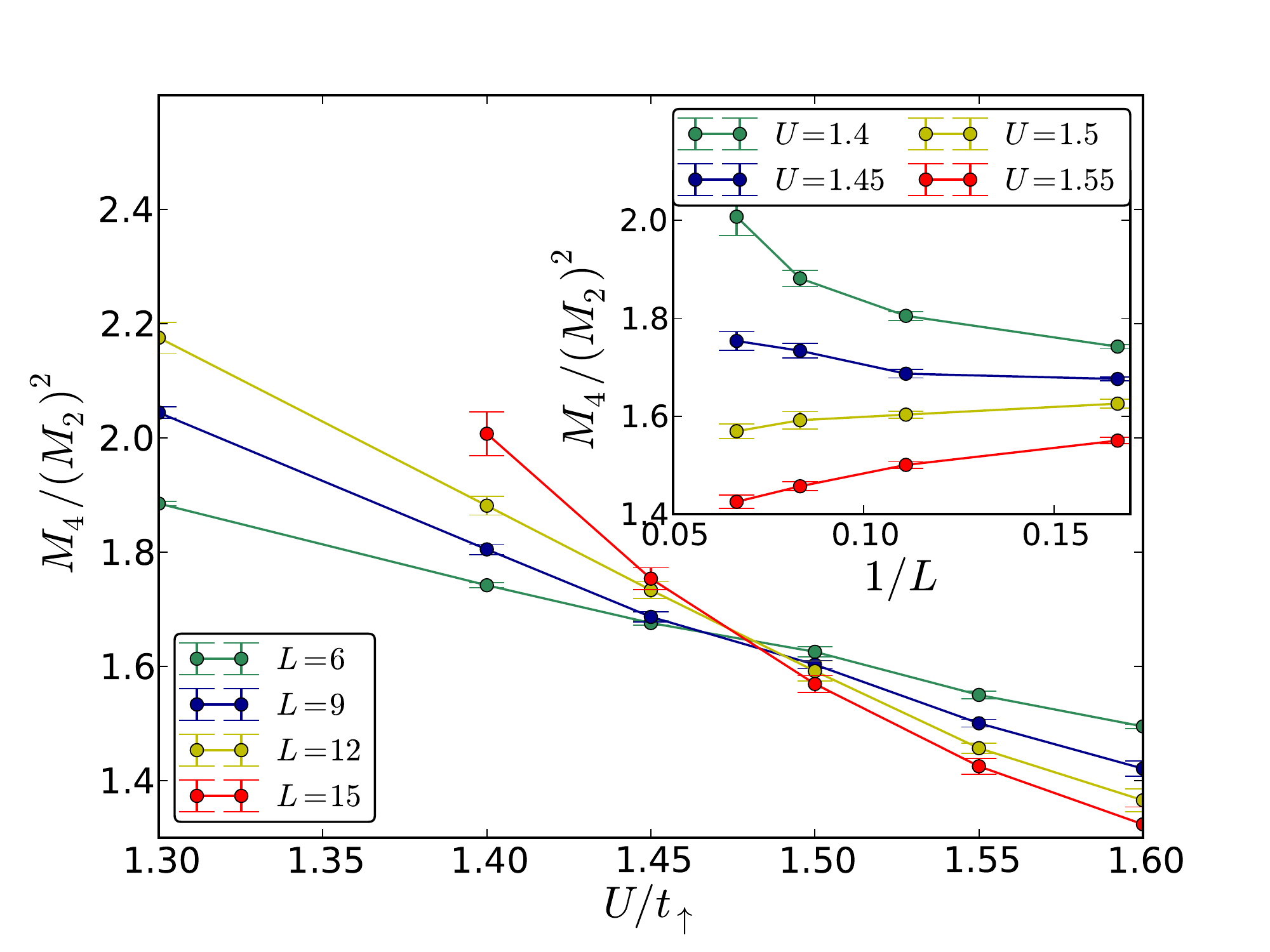}
\caption{Binder ratios calculated at ground state for the honeycomb lattice with $t_\downarrow/t_\uparrow=0.15$. Inset shows the same data plotted versus inverse system length. The quantum critical point is estimated between $U/t_{\uparrow}=1.45\sim 1.5$.
\label{fig:Binder}} 
\end{figure}

Figure~\ref{fig:Binder} shows the Binder ratio for various system sizes. The crossing point suggests a critical value between $U/t_{\uparrow}=1.45\sim1.5$, substantially smaller than the critical point of the $SU(2)$ case $U/t_{\uparrow} \approx 3.8$.\cite{Sorella:2012hib,Assaad:2013kg} This value also differs from a simple renormalization group estimate described in Appendix~\ref{appendix:rg}. Inset of Fig.~\ref{fig:Binder} shows the Binder ratio versus the inverse system lengths, where a size independent value between $U/t_{\uparrow}=1.45\sim1.5$ is clearly visible. Besides, the value of the Binder ratio crossing is quite far from the universal value of the Ising phase transition, suggesting a different universal class. 

\begin{figure}[t]
\includegraphics[width=8cm]{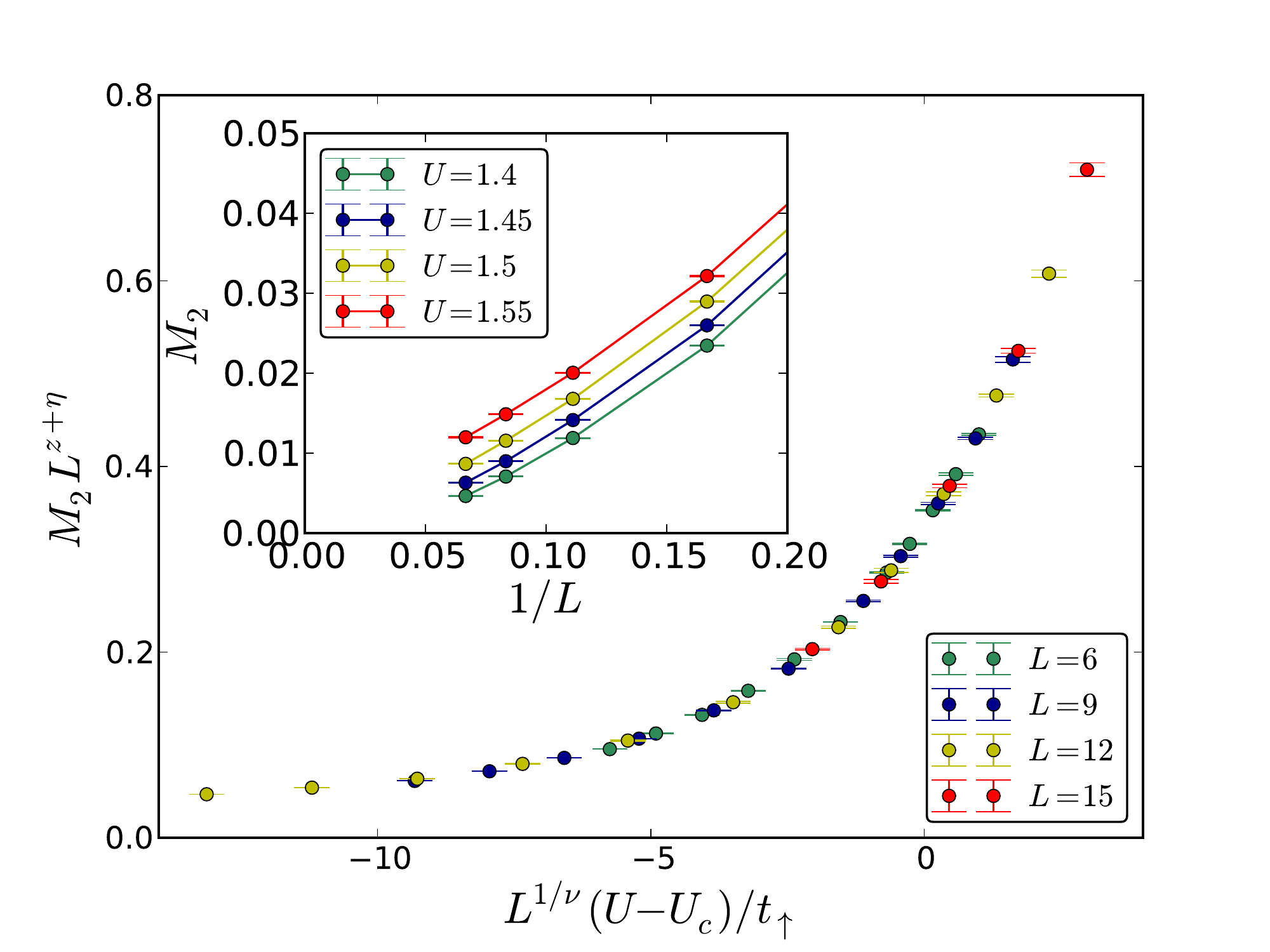}
\caption{
Staggered spin structure factor Eq.~(\ref{eq:M2}) on the honeycomb lattice, scaled according to \Eq{eq:fss} with $U_{c}/t_{\uparrow}=1.481$, $\nu = 0.84$, and $z+\eta = 1.395$.  Inset shows the same data (unscaled) plotted against the inverse system length.  
\label{fig:M2}} 
\end{figure}

We proceed to estimate the critical exponents of this quantum phase transition. Close to the quantum critical point the spin structure factor follows the scaling ansatz,  
\begin{equation}
	M_{2}= L^{-z-\eta} \mathcal{F}\left[L^{1/\nu}\left(U-U_{c}\right)\right], \label{eq:fss}
\end{equation}
where $\mathcal{F}$ is a universal function, $\nu$ is the correlation length critical exponent, $z$ is the dynamical critical exponent, $\eta$ is the anomalous dimension. Fitting to \Eq{eq:fss} gives estimates for the critical point $U_{c}/t_{\uparrow}=1.481(2)$, and critical exponents $\nu=0.84(4)$ and $z+\eta=1.395(7)$. Figure~\ref{fig:M2} shows an excellent  collapse of the scaled $M_{2}$ data using these values. As expected, the critical exponents are  different from the ones of the $SU(2)$ symmetric Hubbard model $\nu=0.88$ and $z+\eta=1.8$\cite{Herbut:2009gaa, Assaad:2013kg} because of the reduced symmetry of the model~(\ref{eq:Ham}).\footnote{The critical exponents are also different from the ones obtained for spinless fermions on honeycomb and $\pi$ flux lattices\cite{Wang:2014iba, Wang:2015tf, Li:2015cwb} possibly due to a different number of fermion flavors.} The inset of Fig.~\ref{fig:M2} shows the structure factor versus inverse system length, which should converge to the square of the antiferromagnetic order parameter as the system size grows. As is emphasized in~\Ref{Sandvik:2012fka}, determining the critical point solely from the $1/L$-extrapolation of structure factors calculated for limited system sizes is  difficult.\cite{Meng:2010gc, Sorella:2012hib} Our experience  also suggests that it is more reliable to extract the critical points from the dimensionless ratios\cite{ParisenToldin:2015gs} such as the Binder ratio in Fig.~\ref{fig:Binder} and the finite size scaling analysis of Fig.~\ref{fig:M2}. 


\section{Summary \label{sec:summary}}
We have presented two sign-problem free CT-QMC methods for efficient simulation of mass-imbalanced Hubbard models~(\ref{eq:Ham}) on bipartite lattices at half-filling. Using them we obtained unbiased results for spin-spin correlations on a one-dimensional chain in Sec.~\ref{sec:1dchain} and the transition temperature to the antiferromagnetic Ising state on the square lattice in Sec.~\ref{sec:square}. These predictions are relevant to the ongoing experimental efforts of observing magnetic long-range orders in ultracold fermion systems.\cite{Greif:2013kb,Hart:2015exa}
We also determine the location and critical exponents of the quantum phase transition of model~(\ref{eq:Ham}) on the honeycomb lattice in Sec.~\ref{sec:honeycomb}, which is relevant for studies of novel fermionic quantum criticality of Dirac fermions.\cite{Herbut:2006jaa, Sorella:2012hib, Assaad:2013kg, ParisenToldin:2015gs}  

These developments open the door to answer several open  problems. Reference~\onlinecite{PhysRevLett.109.065301} reported on the advantage of achieving magnetic long-range orders in mass-imbalanced fermion systems based on approximate DMFT calculations. Now, using the methods developed in this paper, it is possible to unbiasedly examine the magnetic and thermodynamic properties of the mass-imbalanced Hubbard model on a three-dimensional cubic lattice to provide quantitative guides for experimental efforts. On the other hand, the observed fermionic quantum critical point next to an Ising symmetry broken phase in the honeycomb lattice provides yet another opportunity to crosscheck the field theory predictions based on the Gross-Neveu model.\cite{Herbut:2006jaa} It is also interesting to examine the phase diagrams and crossovers sketched in Fig.~\ref{fig:phasediag} for general hopping asymmetries. Furthermore, our methods can be readily applied to Hubbard models with arbitrary spin-dependent anisotropic hoppings, which are relevant for multi-orbital systems.\cite{PhysRevLett.113.195301} The method of correlated double-vertex updates and the corresponding measurement scheme can be used in CT-INT simulations of cluster models in the context of dynamical cluster approximation calculations.\cite{Shinaoka:2015wq}

\acknowledgements
We thank Rama Chitra, R\'emi Desbuquois, Daniel Greif, Igor Herbut, Mauro Iazzi, Jakub Imri\v{s}ka, Gregor Jotzu, Evgeny Kozik, Michael Messer, Maurice Rice, Hiroshi Shinaoka, Matthias Troyer and Yang Zhang for helpful discussions. We are also grateful to Jakub Imri\v{s}ka, Maurice Rice and Matthias Troyer  for their careful reading of the manuscript. Simulations were performed on the M\"{o}nch  and Brutus cluster of ETH Zurich. We have used ALPS libraries\cite{BBauer:2011tz} for Monte Carlo simulations and data analysis. This work was supported by ERC Advanced Grant SIMCOFE and by the Swiss National Science Foundation through the National Center of Competence in Research Quantum Science and Technology QSIT.

\appendix

\section{Measurement of the transverse spin correlation in CT-INT \label{appendix:inplane}}

In terms of fermion operators, the spin correlation in $x$ direction is 
\begin{equation}
\braket{\hat{S}_{\mathbf{i}}^{x}\hat{S}_{\mathbf{j}}^{x}}=\begin{cases} -\frac{1}{4}\braket{\hat{c}_{\mathbf{i}\uparrow}\hat{c}_{\mathbf{j}\uparrow}^{\dagger}\hat{c}_{\mathbf{j}\downarrow}\hat{c}_{\mathbf{i}\downarrow}^{\dagger}+\mbox{h.c.}}  & \mbox{if } \mathbf{i}\neq \mathbf{j}, \\
-\frac{1}{2}\left\langle\hat{n}_{\mathbf{i}\uparrow}\hat{n}_{\mathbf{i}\downarrow}\right\rangle+\frac{1}{4}\braket{\hat{n}_{\mathbf{i}\uparrow}+\hat{n}_{\mathbf{i}\downarrow}} & \mbox{if } \mathbf{i} =  \mathbf{j}, \end{cases} 
\label{eq:xopt}
\end{equation}
where the two terms in the first line are equal because the Hamiltonian is real. 

Similarly to the longitudinal case, we expand the observable as follows
\begin{eqnarray}
\braket{\hat{c}_{\mathbf{i}\uparrow}\hat{c}_{\mathbf{j}\uparrow}^{\dagger}\hat{c}_{\mathbf{j}\downarrow}\hat{c}_{\mathbf{i}\downarrow}^{\dagger}} 
&=&\frac{Z_{0}}{Z}\sum_{k=0}^{\infty} \left(-U\right)^{k}\sum_{\mathcal{C}_{k}}   \nonumber\\
&&\times \det G^{\uparrow  }\left(\mathcal{C}_{k}\cup(\mathbf{i,j},\tau)\right)   \nonumber \\
&&\times \det G^{\downarrow}\left(\mathcal{C}_{k}\cup(\mathbf{j,i},\tau)\right),  \label{eq:spinflip}
\end{eqnarray}
where $G^{\sigma}\left(\mathcal{C}_{k}\cup(\mathbf{i,j},\tau)\right)$ is obtained by inserting one row and one column to $G^{\sigma}\left(\mathcal{C}_{k}\right)$, 
\begin{equation}
\newcommand*{\temp}{\multicolumn{1}{r|}{}} 
G^{\sigma}\left(\mathcal{C}_{k}\cup(\mathbf{i,j},\tau)\right)=\left(\begin{array}{ccc}
G^{\sigma}\left(\mathcal{C}_{k}\right) &\temp &  \mathcal{G}^{\sigma}_{\mathbf{i}_{p} \mathbf{j}}(\tau_{p}-\tau)  \\ \cline{1-3} 
\mathcal{G}^{\sigma }_{\mathbf{i} \mathbf{i}_{q}}(\tau-\tau_{q})   &\temp& \mathcal{G}^{\sigma}_{\mathbf{i} \mathbf{j}}(0^{+})
\end{array}\right).
\end{equation}
The contribution to \Eq{eq:spinflip} is nonzero for both even and odd expansion orders. The estimator for even $k$ reads
\begin{eqnarray}\braket{\hat{c}_{\mathbf{i}\uparrow}\hat{c}_{\mathbf{j}\uparrow}^{\dagger}\hat{c}_{\mathbf{j}\downarrow}\hat{c}_{\mathbf{i}\downarrow}^{\dagger}}_{\mathcal{C}_{k},\tau}^{(1)} &=&  
\frac{\det G^{\uparrow}\left(\mathcal{C}_{k}\cup(\mathbf{i,j},\tau)\right)}{\det G^{\uparrow}\left(\mathcal{C}_{k}\right)} \nonumber\\
&\times& \frac{\det G^{\downarrow}\left(\mathcal{C}_{k}\cup(\mathbf{j,i},\tau)\right)}{\det G^{\downarrow}\left(\mathcal{C}_{k}\right)}. \label{eq:part1}
\end{eqnarray}
For odd $k$, there is again the \emph{zero times infinity} contribution to the observable, which is obtained by the shift measurement. 
The corresponding estimator is
\begin{eqnarray}
&&\braket{ \hat{c}_{\mathbf{i}\uparrow}\hat{c}_{\mathbf{j}\uparrow}^{\dagger}\hat{c}_{\mathbf{j}\downarrow}\hat{c}_{\mathbf{i}\downarrow}^{\dagger}}_{\mathcal{C}_{k},\tau_{p} }^{(2)}  =  \frac{-k}{U\beta N} \label{eq:part2} \\
&\times &\frac{\det G^{\uparrow}\left(\mathcal{C}_{k}\setminus(\mathbf{i}_{p},\tau_{p})\cup(\mathbf{i}_{p}, \mathbf{i}_{p}+\mathbf{j}-\mathbf{i},\tau_{p}) \right)}{\det G^{\uparrow}\left(\mathcal{C}_{k}\right)}\nonumber\\
&\times &\frac{\det G^{\downarrow}\left(\mathcal{C}_{k}\setminus(\mathbf{i}_{p},\tau_{p})\cup(\mathbf{i}_{p}+\mathbf{j}-\mathbf{i},\mathbf{i}_{p},\tau_{p}) \right)}{\det G^{\downarrow}\left(\mathcal{C}_{k}\right)} \nonumber, 
\end{eqnarray} 
where $G^{\sigma}\left(\mathcal{C}_{k}\setminus(\mathbf{i}_{p},\tau_{p})\cup(\mathbf{i}_{p},\mathbf{i}_{p}+\mathbf{j}-\mathbf{i},\tau_{p}) \right)$ is obtained from $G^{\sigma}\left(\mathcal{C}_{k}\right)$ by changing the $p$th row and column. Here the shifted matrix has a nonzero diagonal element $G^\sigma_{pp}$, however we could still use \Eq{eq:shift_ratio} to calculate the determinant ratio because the derivation does not rely on $S=0$. The correct result is obtained by summing the two contributions \Eq{eq:part1} and \Eq{eq:part2}. 

\section{Renormalization group analysis of the quantum phase transition on honeycomb lattice \label{appendix:rg}}
Analytically, the quantum phase transition in the symmetric case was studied by the one-loop Wilson renormalization group (RG) using a large-$N$ expansion.\cite{Herbut:2006jaa} Here $N$ means the number of fermion species, and physically $N=2$ for the two spins. In the field theoretical treatment, the Hubbard $U$ is mapped to a four-fermion coupling constant $g$, which is proportional to $U$. The $\beta$-function for $g$ is $\beta\left(g\right)=\beta_{1}g+\beta_{2}g^{2}+O(g^{3},g^2/N)$.\cite{Herbut:2006jaa} The first coefficient $\beta_{1}=-1$ is derived by power counting, and thus not affected by the hopping asymmetry. The second coefficient $\beta_{2}$ is a one-loop correction. In the large-$N$ limit, only Feynman diagrams with the largest number of fermion loops contribute. For the four-fermion interaction it is one single particle-hole bubble. Thus $\beta_{2}$ is proportional to
\begin{equation}
	\beta_{2}\propto\sum_{\sigma}\int_{-\infty}^{\infty}\frac{d\omega}{2\pi}\int_{\left|\mathbf{k}\right|=\Lambda/b}^{\Lambda}\frac{d^{2}k}{\left(2\pi\right)^{2}}\frac{1}{\omega^{2}+v_{\sigma}^{2}k^{2}}. 
\end{equation}
Here $b>1$ is the scaling factor, $\Lambda$ is the physical cutoff (or inverse lattice spacing), $v_{\sigma}\propto t_{\sigma}$ is the Fermi velocity for spin $\sigma$ at the Dirac cone. Making the substitution $\omega\rightarrow v_{\sigma}\omega$, one finds $\beta_{2}\propto\sum_{\sigma}t_{\sigma}^{-1}$. The quantum critical point, which is the solution of $\beta\left(g_{c}\right)=0$, depends on the hopping asymmetry as follows 
\begin{eqnarray} 
g_{c} &\propto& \beta_{2}^{-1}\propto	U_{c}  \propto \frac{1}{1/t_{\uparrow}+1/{t_{\downarrow}}}, \\
U_{c}  &\approx& \frac{7.6t_\uparrow}{1+t_\uparrow/t_\downarrow},
\label{eq:rg}
\end{eqnarray}
where in the last step we used the known result for the symmetric case $U_c(t_\uparrow,t_\uparrow)\approx 3.8t_\uparrow$.\cite{Sorella:2012hib,Assaad:2013kg}  Note that RG correctly reproduces the exact result $U_{c}(t_\uparrow,0)=0$ in the Falicov-Kimball limit. For the case $t_\downarrow/t_\uparrow=0.15$, it predicts $U_c(t_\uparrow,0.15t_\uparrow)/t_\uparrow\approx7.6/(1+1/0.15)=0.99$, which is significantly smaller than the CT-QMC result $U_c(t_\uparrow,0.15t_\uparrow)/t_\uparrow = 1.45 \sim 1.5$. The approximations made in the RG analysis might break down in the general mass-imbalanced case. This calls for further developments in RG analysis. 

\bibliographystyle{apsrev4-1} 
\bibliography{AHM}

\begin{thebibliography}{71}%
\makeatletter
\providecommand \@ifxundefined [1]{%
 \@ifx{#1\undefined}
}%
\providecommand \@ifnum [1]{%
 \ifnum #1\expandafter \@firstoftwo
 \else \expandafter \@secondoftwo
 \fi
}%
\providecommand \@ifx [1]{%
 \ifx #1\expandafter \@firstoftwo
 \else \expandafter \@secondoftwo
 \fi
}%
\providecommand \natexlab [1]{#1}%
\providecommand \enquote  [1]{``#1''}%
\providecommand \bibnamefont  [1]{#1}%
\providecommand \bibfnamefont [1]{#1}%
\providecommand \citenamefont [1]{#1}%
\providecommand \href@noop [0]{\@secondoftwo}%
\providecommand \href [0]{\begingroup \@sanitize@url \@href}%
\providecommand \@href[1]{\@@startlink{#1}\@@href}%
\providecommand \@@href[1]{\endgroup#1\@@endlink}%
\providecommand \@sanitize@url [0]{\catcode `\\12\catcode `\$12\catcode
  `\&12\catcode `\#12\catcode `\^12\catcode `\_12\catcode `\%12\relax}%
\providecommand \@@startlink[1]{}%
\providecommand \@@endlink[0]{}%
\providecommand \url  [0]{\begingroup\@sanitize@url \@url }%
\providecommand \@url [1]{\endgroup\@href {#1}{\urlprefix }}%
\providecommand \urlprefix  [0]{URL }%
\providecommand \Eprint [0]{\href }%
\providecommand \doibase [0]{http://dx.doi.org/}%
\providecommand \selectlanguage [0]{\@gobble}%
\providecommand \bibinfo  [0]{\@secondoftwo}%
\providecommand \bibfield  [0]{\@secondoftwo}%
\providecommand \translation [1]{[#1]}%
\providecommand \BibitemOpen [0]{}%
\providecommand \bibitemStop [0]{}%
\providecommand \bibitemNoStop [0]{.\EOS\space}%
\providecommand \EOS [0]{\spacefactor3000\relax}%
\providecommand \BibitemShut  [1]{\csname bibitem#1\endcsname}%
\let\auto@bib@innerbib\@empty
\bibitem [{\citenamefont {Huffman}\ and\ \citenamefont
  {Chandrasekharan}(2014)}]{Huffman:2014fj}%
  \BibitemOpen
  \bibfield  {author} {\bibinfo {author} {\bibfnamefont {E.~F.}\ \bibnamefont
  {Huffman}}\ and\ \bibinfo {author} {\bibfnamefont {S.}~\bibnamefont
  {Chandrasekharan}},\ }\href
  {http://link.aps.org/doi/10.1103/PhysRevB.89.111101} {\bibfield  {journal}
  {\bibinfo  {journal} {Phys. Rev. B}\ }\textbf {\bibinfo {volume} {89}},\
  \bibinfo {pages} {111101} (\bibinfo {year} {2014})}\BibitemShut {NoStop}%
\bibitem [{\citenamefont {Li}\ \emph {et~al.}(2015{\natexlab{a}})\citenamefont
  {Li}, \citenamefont {Jiang},\ and\ \citenamefont {Yao}}]{Li:2015jf}%
  \BibitemOpen
  \bibfield  {author} {\bibinfo {author} {\bibfnamefont {Z.-X.}\ \bibnamefont
  {Li}}, \bibinfo {author} {\bibfnamefont {Y.-F.}\ \bibnamefont {Jiang}}, \
  and\ \bibinfo {author} {\bibfnamefont {H.}~\bibnamefont {Yao}},\ }\href
  {http://link.aps.org/doi/10.1103/PhysRevB.91.241117} {\bibfield  {journal}
  {\bibinfo  {journal} {Phys. Rev. B}\ }\textbf {\bibinfo {volume} {91}},\
  \bibinfo {pages} {241117} (\bibinfo {year} {2015}{\natexlab{a}})}\BibitemShut
  {NoStop}%
\bibitem [{\citenamefont {Wang}\ \emph
  {et~al.}(2015{\natexlab{a}})\citenamefont {Wang}, \citenamefont {Liu},
  \citenamefont {Iazzi}, \citenamefont {Troyer},\ and\ \citenamefont
  {Harcos}}]{Wang:2015vw}%
  \BibitemOpen
  \bibfield  {author} {\bibinfo {author} {\bibfnamefont {L.}~\bibnamefont
  {Wang}}, \bibinfo {author} {\bibfnamefont {Y.-H.}\ \bibnamefont {Liu}},
  \bibinfo {author} {\bibfnamefont {M.}~\bibnamefont {Iazzi}}, \bibinfo
  {author} {\bibfnamefont {M.}~\bibnamefont {Troyer}}, \ and\ \bibinfo {author}
  {\bibfnamefont {G.}~\bibnamefont {Harcos}},\ }\href
  {http://arxiv.org/abs/1506.05349v1} {\bibfield  {journal} {\bibinfo
  {journal} {arXiv:1506.05349}\ } (\bibinfo {year}
  {2015}{\natexlab{a}})}\BibitemShut {NoStop}%
\bibitem [{\citenamefont {Iazzi}\ and\ \citenamefont
  {Troyer}(2015)}]{Iazzi:2014vv}%
  \BibitemOpen
  \bibfield  {author} {\bibinfo {author} {\bibfnamefont {M.}~\bibnamefont
  {Iazzi}}\ and\ \bibinfo {author} {\bibfnamefont {M.}~\bibnamefont {Troyer}},\
  }\href {\doibase 10.1103/PhysRevB.91.241118} {\bibfield  {journal} {\bibinfo
  {journal} {Phys. Rev. B}\ }\textbf {\bibinfo {volume} {91}},\ \bibinfo
  {pages} {241118} (\bibinfo {year} {2015})}\BibitemShut {NoStop}%
\bibitem [{\citenamefont {Wang}\ \emph
  {et~al.}(2015{\natexlab{b}})\citenamefont {Wang}, \citenamefont {Iazzi},
  \citenamefont {Corboz},\ and\ \citenamefont {Troyer}}]{Wang:2015tf}%
  \BibitemOpen
  \bibfield  {author} {\bibinfo {author} {\bibfnamefont {L.}~\bibnamefont
  {Wang}}, \bibinfo {author} {\bibfnamefont {M.}~\bibnamefont {Iazzi}},
  \bibinfo {author} {\bibfnamefont {P.}~\bibnamefont {Corboz}}, \ and\ \bibinfo
  {author} {\bibfnamefont {M.}~\bibnamefont {Troyer}},\ }\href {\doibase
  10.1103/PhysRevB.91.235151} {\bibfield  {journal} {\bibinfo  {journal} {Phys.
  Rev. B}\ }\textbf {\bibinfo {volume} {91}},\ \bibinfo {pages} {235151}
  (\bibinfo {year} {2015}{\natexlab{b}})}\BibitemShut {NoStop}%
\bibitem [{\citenamefont {Scalapino}\ \emph {et~al.}(1984)\citenamefont
  {Scalapino}, \citenamefont {Sugar},\ and\ \citenamefont
  {Toussaint}}]{Scalapino:1984wz}%
  \BibitemOpen
  \bibfield  {author} {\bibinfo {author} {\bibfnamefont {D.~J.}\ \bibnamefont
  {Scalapino}}, \bibinfo {author} {\bibfnamefont {R.~L.}\ \bibnamefont
  {Sugar}}, \ and\ \bibinfo {author} {\bibfnamefont {W.~D.}\ \bibnamefont
  {Toussaint}},\ }\href {http://prb.aps.org/abstract/PRB/v29/i9/p5253_1}
  {\bibfield  {journal} {\bibinfo  {journal} {Phys. Rev. B}\ }\textbf {\bibinfo
  {volume} {29}},\ \bibinfo {pages} {5253} (\bibinfo {year}
  {1984})}\BibitemShut {NoStop}%
\bibitem [{\citenamefont {Gubernatis}\ \emph {et~al.}(1985)\citenamefont
  {Gubernatis}, \citenamefont {Scalapino}, \citenamefont {Sugar},\ and\
  \citenamefont {Toussaint}}]{Gubernatis:1985wo}%
  \BibitemOpen
  \bibfield  {author} {\bibinfo {author} {\bibfnamefont {J.~E.}\ \bibnamefont
  {Gubernatis}}, \bibinfo {author} {\bibfnamefont {D.~J.}\ \bibnamefont
  {Scalapino}}, \bibinfo {author} {\bibfnamefont {R.~L.}\ \bibnamefont
  {Sugar}}, \ and\ \bibinfo {author} {\bibfnamefont {W.~D.}\ \bibnamefont
  {Toussaint}},\ }\href {http://prb.aps.org/abstract/PRB/v32/i1/p103_1}
  {\bibfield  {journal} {\bibinfo  {journal} {Phys. Rev. B}\ }\textbf {\bibinfo
  {volume} {32}},\ \bibinfo {pages} {103} (\bibinfo {year} {1985})}\BibitemShut
  {NoStop}%
\bibitem [{\citenamefont {Chandrasekharan}\ and\ \citenamefont
  {Wiese}(1999)}]{Anonymous:xaWVK-gC}%
  \BibitemOpen
  \bibfield  {author} {\bibinfo {author} {\bibfnamefont {S.}~\bibnamefont
  {Chandrasekharan}}\ and\ \bibinfo {author} {\bibfnamefont {U.-J.}\
  \bibnamefont {Wiese}},\ }\href
  {http://prl.aps.org/abstract/PRL/v83/i16/p3116_1} {\bibfield  {journal}
  {\bibinfo  {journal} {Phys. Rev. Lett.}\ }\textbf {\bibinfo {volume} {83}},\
  \bibinfo {pages} {3116} (\bibinfo {year} {1999})}\BibitemShut {NoStop}%
\bibitem [{\citenamefont {Wang}\ \emph {et~al.}(2014)\citenamefont {Wang},
  \citenamefont {Corboz},\ and\ \citenamefont {Troyer}}]{Wang:2014iba}%
  \BibitemOpen
  \bibfield  {author} {\bibinfo {author} {\bibfnamefont {L.}~\bibnamefont
  {Wang}}, \bibinfo {author} {\bibfnamefont {P.}~\bibnamefont {Corboz}}, \ and\
  \bibinfo {author} {\bibfnamefont {M.}~\bibnamefont {Troyer}},\ }\href
  {http://iopscience.iop.org/1367-2630/16/10/103008/} {\bibfield  {journal}
  {\bibinfo  {journal} {New J. Phys.}\ }\textbf {\bibinfo {volume} {16}},\
  \bibinfo {pages} {103008} (\bibinfo {year} {2014})}\BibitemShut {NoStop}%
\bibitem [{\citenamefont {Li}\ \emph {et~al.}(2015{\natexlab{b}})\citenamefont
  {Li}, \citenamefont {Jiang},\ and\ \citenamefont {Yao}}]{Li:2015cwb}%
  \BibitemOpen
  \bibfield  {author} {\bibinfo {author} {\bibfnamefont {Z.-X.}\ \bibnamefont
  {Li}}, \bibinfo {author} {\bibfnamefont {Y.-F.}\ \bibnamefont {Jiang}}, \
  and\ \bibinfo {author} {\bibfnamefont {H.}~\bibnamefont {Yao}},\ }\href
  {\doibase 10.1088/1367-2630/17/8/085003} {\bibfield  {journal} {\bibinfo
  {journal} {New J. Phys.}\ }\textbf {\bibinfo {volume} {17}},\ \bibinfo
  {pages} {1} (\bibinfo {year} {2015}{\natexlab{b}})}\BibitemShut {NoStop}%
\bibitem [{Note1()}]{Note1}%
  \BibitemOpen
  \bibinfo {note} {At half-filling the physics of $U<0$ is simply related by a
  particle-hole transformation.}\BibitemShut {Stop}%
\bibitem [{\citenamefont {Taglieber}\ \emph {et~al.}(2008)\citenamefont
  {Taglieber}, \citenamefont {Voigt}, \citenamefont {Aoki}, \citenamefont
  {H\"ansch},\ and\ \citenamefont {Dieckmann}}]{PhysRevLett.100.010401}%
  \BibitemOpen
  \bibfield  {author} {\bibinfo {author} {\bibfnamefont {M.}~\bibnamefont
  {Taglieber}}, \bibinfo {author} {\bibfnamefont {A.-C.}\ \bibnamefont
  {Voigt}}, \bibinfo {author} {\bibfnamefont {T.}~\bibnamefont {Aoki}},
  \bibinfo {author} {\bibfnamefont {T.~W.}\ \bibnamefont {H\"ansch}}, \ and\
  \bibinfo {author} {\bibfnamefont {K.}~\bibnamefont {Dieckmann}},\ }\href
  {\doibase 10.1103/PhysRevLett.100.010401} {\bibfield  {journal} {\bibinfo
  {journal} {Phys. Rev. Lett.}\ }\textbf {\bibinfo {volume} {100}},\ \bibinfo
  {pages} {010401} (\bibinfo {year} {2008})}\BibitemShut {NoStop}%
\bibitem [{\citenamefont {Spiegelhalder}\ \emph {et~al.}(2009)\citenamefont
  {Spiegelhalder}, \citenamefont {Trenkwalder}, \citenamefont {Naik},
  \citenamefont {Hendl}, \citenamefont {Schreck},\ and\ \citenamefont
  {Grimm}}]{PhysRevLett.103.223203}%
  \BibitemOpen
  \bibfield  {author} {\bibinfo {author} {\bibfnamefont {F.~M.}\ \bibnamefont
  {Spiegelhalder}}, \bibinfo {author} {\bibfnamefont {A.}~\bibnamefont
  {Trenkwalder}}, \bibinfo {author} {\bibfnamefont {D.}~\bibnamefont {Naik}},
  \bibinfo {author} {\bibfnamefont {G.}~\bibnamefont {Hendl}}, \bibinfo
  {author} {\bibfnamefont {F.}~\bibnamefont {Schreck}}, \ and\ \bibinfo
  {author} {\bibfnamefont {R.}~\bibnamefont {Grimm}},\ }\href {\doibase
  10.1103/PhysRevLett.103.223203} {\bibfield  {journal} {\bibinfo  {journal}
  {Phys. Rev. Lett.}\ }\textbf {\bibinfo {volume} {103}},\ \bibinfo {pages}
  {223203} (\bibinfo {year} {2009})}\BibitemShut {NoStop}%
\bibitem [{\citenamefont {Tiecke}\ \emph {et~al.}(2010)\citenamefont {Tiecke},
  \citenamefont {Goosen}, \citenamefont {Ludewig}, \citenamefont {Gensemer},
  \citenamefont {Kraft}, \citenamefont {Kokkelmans},\ and\ \citenamefont
  {Walraven}}]{PhysRevLett.104.053202}%
  \BibitemOpen
  \bibfield  {author} {\bibinfo {author} {\bibfnamefont {T.~G.}\ \bibnamefont
  {Tiecke}}, \bibinfo {author} {\bibfnamefont {M.~R.}\ \bibnamefont {Goosen}},
  \bibinfo {author} {\bibfnamefont {A.}~\bibnamefont {Ludewig}}, \bibinfo
  {author} {\bibfnamefont {S.~D.}\ \bibnamefont {Gensemer}}, \bibinfo {author}
  {\bibfnamefont {S.}~\bibnamefont {Kraft}}, \bibinfo {author} {\bibfnamefont
  {S.~J. J. M.~F.}\ \bibnamefont {Kokkelmans}}, \ and\ \bibinfo {author}
  {\bibfnamefont {J.~T.~M.}\ \bibnamefont {Walraven}},\ }\href {\doibase
  10.1103/PhysRevLett.104.053202} {\bibfield  {journal} {\bibinfo  {journal}
  {Phys. Rev. Lett.}\ }\textbf {\bibinfo {volume} {104}},\ \bibinfo {pages}
  {053202} (\bibinfo {year} {2010})}\BibitemShut {NoStop}%
\bibitem [{\citenamefont {Taie}\ \emph {et~al.}(2010)\citenamefont {Taie},
  \citenamefont {Takasu}, \citenamefont {Sugawa}, \citenamefont {Yamazaki},
  \citenamefont {Tsujimoto}, \citenamefont {Murakami},\ and\ \citenamefont
  {Takahashi}}]{PhysRevLett.105.190401}%
  \BibitemOpen
  \bibfield  {author} {\bibinfo {author} {\bibfnamefont {S.}~\bibnamefont
  {Taie}}, \bibinfo {author} {\bibfnamefont {Y.}~\bibnamefont {Takasu}},
  \bibinfo {author} {\bibfnamefont {S.}~\bibnamefont {Sugawa}}, \bibinfo
  {author} {\bibfnamefont {R.}~\bibnamefont {Yamazaki}}, \bibinfo {author}
  {\bibfnamefont {T.}~\bibnamefont {Tsujimoto}}, \bibinfo {author}
  {\bibfnamefont {R.}~\bibnamefont {Murakami}}, \ and\ \bibinfo {author}
  {\bibfnamefont {Y.}~\bibnamefont {Takahashi}},\ }\href {\doibase
  10.1103/PhysRevLett.105.190401} {\bibfield  {journal} {\bibinfo  {journal}
  {Phys. Rev. Lett.}\ }\textbf {\bibinfo {volume} {105}},\ \bibinfo {pages}
  {190401} (\bibinfo {year} {2010})}\BibitemShut {NoStop}%
\bibitem [{\citenamefont {Trenkwalder}\ \emph {et~al.}(2011)\citenamefont
  {Trenkwalder}, \citenamefont {Kohstall}, \citenamefont {Zaccanti},
  \citenamefont {Naik}, \citenamefont {Sidorov}, \citenamefont {Schreck},\ and\
  \citenamefont {Grimm}}]{PhysRevLett.106.115304}%
  \BibitemOpen
  \bibfield  {author} {\bibinfo {author} {\bibfnamefont {A.}~\bibnamefont
  {Trenkwalder}}, \bibinfo {author} {\bibfnamefont {C.}~\bibnamefont
  {Kohstall}}, \bibinfo {author} {\bibfnamefont {M.}~\bibnamefont {Zaccanti}},
  \bibinfo {author} {\bibfnamefont {D.}~\bibnamefont {Naik}}, \bibinfo {author}
  {\bibfnamefont {A.~I.}\ \bibnamefont {Sidorov}}, \bibinfo {author}
  {\bibfnamefont {F.}~\bibnamefont {Schreck}}, \ and\ \bibinfo {author}
  {\bibfnamefont {R.}~\bibnamefont {Grimm}},\ }\href {\doibase
  10.1103/PhysRevLett.106.115304} {\bibfield  {journal} {\bibinfo  {journal}
  {Phys. Rev. Lett.}\ }\textbf {\bibinfo {volume} {106}},\ \bibinfo {pages}
  {115304} (\bibinfo {year} {2011})}\BibitemShut {NoStop}%
\bibitem [{\citenamefont {Kohstall}\ \emph {et~al.}(2013)\citenamefont
  {Kohstall}, \citenamefont {Zaccanti}, \citenamefont {Jag}, \citenamefont
  {Trenkwalder}, \citenamefont {Massignan}, \citenamefont {Bruun},
  \citenamefont {Schreck},\ and\ \citenamefont {Grimm}}]{Kohstall:2013kg}%
  \BibitemOpen
  \bibfield  {author} {\bibinfo {author} {\bibfnamefont {C.}~\bibnamefont
  {Kohstall}}, \bibinfo {author} {\bibfnamefont {M.}~\bibnamefont {Zaccanti}},
  \bibinfo {author} {\bibfnamefont {M.}~\bibnamefont {Jag}}, \bibinfo {author}
  {\bibfnamefont {A.}~\bibnamefont {Trenkwalder}}, \bibinfo {author}
  {\bibfnamefont {P.}~\bibnamefont {Massignan}}, \bibinfo {author}
  {\bibfnamefont {G.~M.}\ \bibnamefont {Bruun}}, \bibinfo {author}
  {\bibfnamefont {F.}~\bibnamefont {Schreck}}, \ and\ \bibinfo {author}
  {\bibfnamefont {R.}~\bibnamefont {Grimm}},\ }\href
  {http://dx.doi.org/10.1038/nature11065} {\bibfield  {journal} {\bibinfo
  {journal} {Nature}\ }\textbf {\bibinfo {volume} {485}},\ \bibinfo {pages}
  {615} (\bibinfo {year} {2013})}\BibitemShut {NoStop}%
\bibitem [{\citenamefont {Jag}\ \emph {et~al.}(2014)\citenamefont {Jag},
  \citenamefont {Zaccanti}, \citenamefont {Cetina}, \citenamefont {Lous},
  \citenamefont {Schreck}, \citenamefont {Grimm}, \citenamefont {Petrov},\ and\
  \citenamefont {Levinsen}}]{Jag:2014gd}%
  \BibitemOpen
  \bibfield  {author} {\bibinfo {author} {\bibfnamefont {M.}~\bibnamefont
  {Jag}}, \bibinfo {author} {\bibfnamefont {M.}~\bibnamefont {Zaccanti}},
  \bibinfo {author} {\bibfnamefont {M.}~\bibnamefont {Cetina}}, \bibinfo
  {author} {\bibfnamefont {R.~S.}\ \bibnamefont {Lous}}, \bibinfo {author}
  {\bibfnamefont {F.}~\bibnamefont {Schreck}}, \bibinfo {author} {\bibfnamefont
  {R.}~\bibnamefont {Grimm}}, \bibinfo {author} {\bibfnamefont {D.~S.}\
  \bibnamefont {Petrov}}, \ and\ \bibinfo {author} {\bibfnamefont
  {J.}~\bibnamefont {Levinsen}},\ }\href
  {http://link.aps.org/doi/10.1103/PhysRevLett.112.075302} {\bibfield
  {journal} {\bibinfo  {journal} {Phys. Rev. Lett.}\ }\textbf {\bibinfo
  {volume} {112}},\ \bibinfo {pages} {075302} (\bibinfo {year}
  {2014})}\BibitemShut {NoStop}%
\bibitem [{\citenamefont {Jotzu}\ \emph {et~al.}(2015)\citenamefont {Jotzu},
  \citenamefont {Messer}, \citenamefont {G\"org}, \citenamefont {Greif},
  \citenamefont {Desbuquois},\ and\ \citenamefont {Esslinger}}]{Jotzu:2015tq}%
  \BibitemOpen
  \bibfield  {author} {\bibinfo {author} {\bibfnamefont {G.}~\bibnamefont
  {Jotzu}}, \bibinfo {author} {\bibfnamefont {M.}~\bibnamefont {Messer}},
  \bibinfo {author} {\bibfnamefont {F.}~\bibnamefont {G\"org}}, \bibinfo
  {author} {\bibfnamefont {D.}~\bibnamefont {Greif}}, \bibinfo {author}
  {\bibfnamefont {R.}~\bibnamefont {Desbuquois}}, \ and\ \bibinfo {author}
  {\bibfnamefont {T.}~\bibnamefont {Esslinger}},\ }\href {\doibase
  10.1103/PhysRevLett.115.073002} {\bibfield  {journal} {\bibinfo  {journal}
  {Phys. Rev. Lett.}\ }\textbf {\bibinfo {volume} {115}},\ \bibinfo {pages}
  {073002} (\bibinfo {year} {2015})}\BibitemShut {NoStop}%
\bibitem [{\citenamefont {Falicov}\ and\ \citenamefont
  {Kimball}(1969)}]{Falicov:1969ua}%
  \BibitemOpen
  \bibfield  {author} {\bibinfo {author} {\bibfnamefont {L.~M.}\ \bibnamefont
  {Falicov}}\ and\ \bibinfo {author} {\bibfnamefont {J.~C.}\ \bibnamefont
  {Kimball}},\ }\href
  {http://journals.aps.org/prl/abstract/10.1103/PhysRevLett.22.997} {\bibfield
  {journal} {\bibinfo  {journal} {Phys. Rev. Lett.}\ }\textbf {\bibinfo
  {volume} {22}},\ \bibinfo {pages} {997} (\bibinfo {year} {1969})}\BibitemShut
  {NoStop}%
\bibitem [{\citenamefont {Brandt}\ and\ \citenamefont
  {Schmidt}(1986)}]{brandt1986exact}%
  \BibitemOpen
  \bibfield  {author} {\bibinfo {author} {\bibfnamefont {U.}~\bibnamefont
  {Brandt}}\ and\ \bibinfo {author} {\bibfnamefont {R.}~\bibnamefont
  {Schmidt}},\ }\href@noop {} {\bibfield  {journal} {\bibinfo  {journal}
  {Zeitschrift f{\"u}r Physik B Condensed Matter}\ }\textbf {\bibinfo {volume}
  {63}},\ \bibinfo {pages} {45} (\bibinfo {year} {1986})}\BibitemShut {NoStop}%
\bibitem [{\citenamefont {Freericks}\ \emph {et~al.}(2002)\citenamefont
  {Freericks}, \citenamefont {Lieb},\ and\ \citenamefont
  {Ueltschi}}]{PhysRevLett.88.106401}%
  \BibitemOpen
  \bibfield  {author} {\bibinfo {author} {\bibfnamefont {J.~K.}\ \bibnamefont
  {Freericks}}, \bibinfo {author} {\bibfnamefont {E.~H.}\ \bibnamefont {Lieb}},
  \ and\ \bibinfo {author} {\bibfnamefont {D.}~\bibnamefont {Ueltschi}},\
  }\href {\doibase 10.1103/PhysRevLett.88.106401} {\bibfield  {journal}
  {\bibinfo  {journal} {Phys. Rev. Lett.}\ }\textbf {\bibinfo {volume} {88}},\
  \bibinfo {pages} {106401} (\bibinfo {year} {2002})}\BibitemShut {NoStop}%
\bibitem [{\citenamefont {Freericks}\ and\ \citenamefont
  {Zlati{\'c}}(2003)}]{2003RvMP...75.1333F}%
  \BibitemOpen
  \bibfield  {author} {\bibinfo {author} {\bibfnamefont {J.~K.}\ \bibnamefont
  {Freericks}}\ and\ \bibinfo {author} {\bibfnamefont {V.}~\bibnamefont
  {Zlati{\'c}}},\ }\href
  {http://adsabs.harvard.edu/cgi-bin/nph-data_query?bibcode=2003RvMP...75.1333F&link_type=ABSTRACT}
  {\bibfield  {journal} {\bibinfo  {journal} {Rev. Mod. Phys.}\ }\textbf
  {\bibinfo {volume} {75}},\ \bibinfo {pages} {1333} (\bibinfo {year}
  {2003})}\BibitemShut {NoStop}%
\bibitem [{\citenamefont {Ma{\'s}ka}\ and\ \citenamefont
  {Czajka}(2006)}]{Maska:2006id}%
  \BibitemOpen
  \bibfield  {author} {\bibinfo {author} {\bibfnamefont {M.}~\bibnamefont
  {Ma{\'s}ka}}\ and\ \bibinfo {author} {\bibfnamefont {K.}~\bibnamefont
  {Czajka}},\ }\href {http://link.aps.org/doi/10.1103/PhysRevB.74.035109}
  {\bibfield  {journal} {\bibinfo  {journal} {Phys. Rev. B}\ }\textbf {\bibinfo
  {volume} {74}},\ \bibinfo {pages} {035109} (\bibinfo {year}
  {2006})}\BibitemShut {NoStop}%
\bibitem [{\citenamefont {{\v Z}onda}\ \emph {et~al.}(2010)\citenamefont {{\v
  Z}onda}, \citenamefont {Farka{\v s}ovsk{\'y}},\ and\ \citenamefont {{\v
  C}en{\v c}arikov{\'a}}}]{Zonda:2010io}%
  \BibitemOpen
  \bibfield  {author} {\bibinfo {author} {\bibfnamefont {M.}~\bibnamefont {{\v
  Z}onda}}, \bibinfo {author} {\bibfnamefont {P.}~\bibnamefont {Farka{\v
  s}ovsk{\'y}}}, \ and\ \bibinfo {author} {\bibfnamefont {H.}~\bibnamefont {{\v
  C}en{\v c}arikov{\'a}}},\ }\href {\doibase 10.1088/1742-6596/200/1/012240}
  {\bibfield  {journal} {\bibinfo  {journal} {J. Phys.: Conf. Ser.}\ }\textbf
  {\bibinfo {volume} {200}},\ \bibinfo {pages} {012240} (\bibinfo {year}
  {2010})}\BibitemShut {NoStop}%
\bibitem [{\citenamefont {Kennedy}\ and\ \citenamefont
  {Lieb}(1986)}]{Kennedy:1986wf}%
  \BibitemOpen
  \bibfield  {author} {\bibinfo {author} {\bibfnamefont {T.}~\bibnamefont
  {Kennedy}}\ and\ \bibinfo {author} {\bibfnamefont {E.~H.}\ \bibnamefont
  {Lieb}},\ }\href
  {http://www.sciencedirect.com/science/article/pii/0378437186901883}
  {\bibfield  {journal} {\bibinfo  {journal} {Physica A}\ }\textbf {\bibinfo
  {volume} {138}},\ \bibinfo {pages} {320} (\bibinfo {year}
  {1986})}\BibitemShut {NoStop}%
\bibitem [{\citenamefont {F\'ath}\ \emph {et~al.}(1995)\citenamefont {F\'ath},
  \citenamefont {Doma\ifmmode~\acute{n}\else \'{n}\fi{}ski},\ and\
  \citenamefont {Lema\ifmmode~\acute{n}\else
  \'{n}\fi{}ski}}]{PhysRevB.52.13910}%
  \BibitemOpen
  \bibfield  {author} {\bibinfo {author} {\bibfnamefont {G.}~\bibnamefont
  {F\'ath}}, \bibinfo {author} {\bibfnamefont {Z.}~\bibnamefont
  {Doma\ifmmode~\acute{n}\else \'{n}\fi{}ski}}, \ and\ \bibinfo {author}
  {\bibfnamefont {R.}~\bibnamefont {Lema\ifmmode~\acute{n}\else
  \'{n}\fi{}ski}},\ }\href {\doibase 10.1103/PhysRevB.52.13910} {\bibfield
  {journal} {\bibinfo  {journal} {Phys. Rev. B}\ }\textbf {\bibinfo {volume}
  {52}},\ \bibinfo {pages} {13910} (\bibinfo {year} {1995})}\BibitemShut
  {NoStop}%
\bibitem [{\citenamefont {Dao}\ \emph {et~al.}(2007)\citenamefont {Dao},
  \citenamefont {Georges},\ and\ \citenamefont {Capone}}]{PhysRevB.76.104517}%
  \BibitemOpen
  \bibfield  {author} {\bibinfo {author} {\bibfnamefont {T.-L.}\ \bibnamefont
  {Dao}}, \bibinfo {author} {\bibfnamefont {A.}~\bibnamefont {Georges}}, \ and\
  \bibinfo {author} {\bibfnamefont {M.}~\bibnamefont {Capone}},\ }\href
  {\doibase 10.1103/PhysRevB.76.104517} {\bibfield  {journal} {\bibinfo
  {journal} {Phys. Rev. B}\ }\textbf {\bibinfo {volume} {76}},\ \bibinfo
  {pages} {104517} (\bibinfo {year} {2007})}\BibitemShut {NoStop}%
\bibitem [{\citenamefont {Dao}\ \emph {et~al.}(2012)\citenamefont {Dao},
  \citenamefont {Ferrero}, \citenamefont {Cornaglia},\ and\ \citenamefont
  {Capone}}]{PhysRevA.85.013606}%
  \BibitemOpen
  \bibfield  {author} {\bibinfo {author} {\bibfnamefont {T.-L.}\ \bibnamefont
  {Dao}}, \bibinfo {author} {\bibfnamefont {M.}~\bibnamefont {Ferrero}},
  \bibinfo {author} {\bibfnamefont {P.~S.}\ \bibnamefont {Cornaglia}}, \ and\
  \bibinfo {author} {\bibfnamefont {M.}~\bibnamefont {Capone}},\ }\href
  {\doibase 10.1103/PhysRevA.85.013606} {\bibfield  {journal} {\bibinfo
  {journal} {Phys. Rev. A}\ }\textbf {\bibinfo {volume} {85}},\ \bibinfo
  {pages} {013606} (\bibinfo {year} {2012})}\BibitemShut {NoStop}%
\bibitem [{\citenamefont {Sotnikov}\ \emph {et~al.}(2012)\citenamefont
  {Sotnikov}, \citenamefont {Cocks},\ and\ \citenamefont
  {Hofstetter}}]{PhysRevLett.109.065301}%
  \BibitemOpen
  \bibfield  {author} {\bibinfo {author} {\bibfnamefont {A.}~\bibnamefont
  {Sotnikov}}, \bibinfo {author} {\bibfnamefont {D.}~\bibnamefont {Cocks}}, \
  and\ \bibinfo {author} {\bibfnamefont {W.}~\bibnamefont {Hofstetter}},\
  }\href {\doibase 10.1103/PhysRevLett.109.065301} {\bibfield  {journal}
  {\bibinfo  {journal} {Phys. Rev. Lett.}\ }\textbf {\bibinfo {volume} {109}},\
  \bibinfo {pages} {065301} (\bibinfo {year} {2012})}\BibitemShut {NoStop}%
\bibitem [{\citenamefont {Fratini}\ and\ \citenamefont
  {Pilati}(2014)}]{PhysRevA.90.023605}%
  \BibitemOpen
  \bibfield  {author} {\bibinfo {author} {\bibfnamefont {E.}~\bibnamefont
  {Fratini}}\ and\ \bibinfo {author} {\bibfnamefont {S.}~\bibnamefont
  {Pilati}},\ }\href {\doibase 10.1103/PhysRevA.90.023605} {\bibfield
  {journal} {\bibinfo  {journal} {Phys. Rev. A}\ }\textbf {\bibinfo {volume}
  {90}},\ \bibinfo {pages} {023605} (\bibinfo {year} {2014})}\BibitemShut
  {NoStop}%
\bibitem [{\citenamefont {Winograd}\ \emph {et~al.}(2011)\citenamefont
  {Winograd}, \citenamefont {Chitra},\ and\ \citenamefont
  {Rozenberg}}]{Winograd:2011en}%
  \BibitemOpen
  \bibfield  {author} {\bibinfo {author} {\bibfnamefont {E.~A.}\ \bibnamefont
  {Winograd}}, \bibinfo {author} {\bibfnamefont {R.}~\bibnamefont {Chitra}}, \
  and\ \bibinfo {author} {\bibfnamefont {M.~J.}\ \bibnamefont {Rozenberg}},\
  }\href {\doibase 10.1103/PhysRevB.84.233102} {\bibfield  {journal} {\bibinfo
  {journal} {Phys. Rev. B}\ }\textbf {\bibinfo {volume} {84}},\ \bibinfo
  {pages} {233102} (\bibinfo {year} {2011})}\BibitemShut {NoStop}%
\bibitem [{\citenamefont {Winograd}\ \emph {et~al.}(2012)\citenamefont
  {Winograd}, \citenamefont {Chitra},\ and\ \citenamefont
  {Rozenberg}}]{Winograd:2012ie}%
  \BibitemOpen
  \bibfield  {author} {\bibinfo {author} {\bibfnamefont {E.~A.}\ \bibnamefont
  {Winograd}}, \bibinfo {author} {\bibfnamefont {R.}~\bibnamefont {Chitra}}, \
  and\ \bibinfo {author} {\bibfnamefont {M.~J.}\ \bibnamefont {Rozenberg}},\
  }\href {\doibase 10.1103/PhysRevB.86.195118} {\bibfield  {journal} {\bibinfo
  {journal} {Phys. Rev. B}\ }\textbf {\bibinfo {volume} {86}},\ \bibinfo
  {pages} {195118} (\bibinfo {year} {2012})}\BibitemShut {NoStop}%
\bibitem [{\citenamefont {Farka\ifmmode~\check{s}\else
  \v{s}\fi{}ovsk\'y}(2008)}]{PhysRevB.77.085110}%
  \BibitemOpen
  \bibfield  {author} {\bibinfo {author} {\bibfnamefont {P.}~\bibnamefont
  {Farka\ifmmode~\check{s}\else \v{s}\fi{}ovsk\'y}},\ }\href {\doibase
  10.1103/PhysRevB.77.085110} {\bibfield  {journal} {\bibinfo  {journal} {Phys.
  Rev. B}\ }\textbf {\bibinfo {volume} {77}},\ \bibinfo {pages} {085110}
  (\bibinfo {year} {2008})}\BibitemShut {NoStop}%
\bibitem [{\citenamefont {Roscher}\ \emph {et~al.}(2014)\citenamefont
  {Roscher}, \citenamefont {Braun}, \citenamefont {Chen},\ and\ \citenamefont
  {Drut}}]{Roscher:2014hb}%
  \BibitemOpen
  \bibfield  {author} {\bibinfo {author} {\bibfnamefont {D.}~\bibnamefont
  {Roscher}}, \bibinfo {author} {\bibfnamefont {J.}~\bibnamefont {Braun}},
  \bibinfo {author} {\bibfnamefont {J.-W.}\ \bibnamefont {Chen}}, \ and\
  \bibinfo {author} {\bibfnamefont {J.~E.}\ \bibnamefont {Drut}},\ }\href
  {http://stacks.iop.org/0954-3899/41/i=5/a=055110?key=crossref.1603f23e46643146532514fce54846d9}
  {\bibfield  {journal} {\bibinfo  {journal} {J. Phys. G: Nucl. Part. Phys.}\
  }\textbf {\bibinfo {volume} {41}},\ \bibinfo {pages} {055110} (\bibinfo
  {year} {2014})}\BibitemShut {NoStop}%
\bibitem [{\citenamefont {Braun}\ \emph {et~al.}(2015)\citenamefont {Braun},
  \citenamefont {Drut},\ and\ \citenamefont {Roscher}}]{Braun:2015du}%
  \BibitemOpen
  \bibfield  {author} {\bibinfo {author} {\bibfnamefont {J.}~\bibnamefont
  {Braun}}, \bibinfo {author} {\bibfnamefont {J.~E.}\ \bibnamefont {Drut}}, \
  and\ \bibinfo {author} {\bibfnamefont {D.}~\bibnamefont {Roscher}},\ }\href
  {http://link.aps.org/doi/10.1103/PhysRevLett.114.050404} {\bibfield
  {journal} {\bibinfo  {journal} {Phys. Rev. Lett.}\ }\textbf {\bibinfo
  {volume} {114}},\ \bibinfo {pages} {050404} (\bibinfo {year}
  {2015})}\BibitemShut {NoStop}%
\bibitem [{\citenamefont {Gezerlis}\ \emph {et~al.}(2009)\citenamefont
  {Gezerlis}, \citenamefont {Gandolfi}, \citenamefont {Schmidt},\ and\
  \citenamefont {Carlson}}]{Gezerlis:2009fb}%
  \BibitemOpen
  \bibfield  {author} {\bibinfo {author} {\bibfnamefont {A.}~\bibnamefont
  {Gezerlis}}, \bibinfo {author} {\bibfnamefont {S.}~\bibnamefont {Gandolfi}},
  \bibinfo {author} {\bibfnamefont {K.~E.}\ \bibnamefont {Schmidt}}, \ and\
  \bibinfo {author} {\bibfnamefont {J.}~\bibnamefont {Carlson}},\ }\href
  {http://link.aps.org/doi/10.1103/PhysRevLett.103.060403} {\bibfield
  {journal} {\bibinfo  {journal} {Phys. Rev. Lett.}\ }\textbf {\bibinfo
  {volume} {103}},\ \bibinfo {pages} {060403} (\bibinfo {year}
  {2009})}\BibitemShut {NoStop}%
\bibitem [{\citenamefont {Kroiss}\ and\ \citenamefont
  {Pollet}(2015)}]{Kroiss:2015dy}%
  \BibitemOpen
  \bibfield  {author} {\bibinfo {author} {\bibfnamefont {P.}~\bibnamefont
  {Kroiss}}\ and\ \bibinfo {author} {\bibfnamefont {L.}~\bibnamefont
  {Pollet}},\ }\href {http://link.aps.org/doi/10.1103/PhysRevB.91.144507}
  {\bibfield  {journal} {\bibinfo  {journal} {Phys. Rev. B}\ }\textbf {\bibinfo
  {volume} {91}},\ \bibinfo {pages} {144507} (\bibinfo {year}
  {2015})}\BibitemShut {NoStop}%
\bibitem [{\citenamefont {S{\"o}yler}\ \emph {et~al.}(2009)\citenamefont
  {S{\"o}yler}, \citenamefont {Capogrosso-Sansone}, \citenamefont {Prokof'ev},\
  and\ \citenamefont {Svistunov}}]{Soyler:2009eya}%
  \BibitemOpen
  \bibfield  {author} {\bibinfo {author} {\bibfnamefont {{\c S}.~G.}\
  \bibnamefont {S{\"o}yler}}, \bibinfo {author} {\bibfnamefont
  {B.}~\bibnamefont {Capogrosso-Sansone}}, \bibinfo {author} {\bibfnamefont
  {N.~V.}\ \bibnamefont {Prokof'ev}}, \ and\ \bibinfo {author} {\bibfnamefont
  {B.~V.}\ \bibnamefont {Svistunov}},\ }\href {\doibase
  10.1088/1367-2630/11/7/073036} {\bibfield  {journal} {\bibinfo  {journal}
  {New J. Phys.}\ }\textbf {\bibinfo {volume} {11}},\ \bibinfo {pages} {073036}
  (\bibinfo {year} {2009})}\BibitemShut {NoStop}%
\bibitem [{\citenamefont {Capogrosso-Sansone}\ \emph
  {et~al.}(2010)\citenamefont {Capogrosso-Sansone}, \citenamefont {S\"oyler},
  \citenamefont {Prokof'ev},\ and\ \citenamefont
  {Svistunov}}]{PhysRevA.81.053622}%
  \BibitemOpen
  \bibfield  {author} {\bibinfo {author} {\bibfnamefont {B.}~\bibnamefont
  {Capogrosso-Sansone}}, \bibinfo {author} {\bibfnamefont {i.~m. c.~G.}\
  \bibnamefont {S\"oyler}}, \bibinfo {author} {\bibfnamefont {N.~V.}\
  \bibnamefont {Prokof'ev}}, \ and\ \bibinfo {author} {\bibfnamefont {B.~V.}\
  \bibnamefont {Svistunov}},\ }\href {\doibase 10.1103/PhysRevA.81.053622}
  {\bibfield  {journal} {\bibinfo  {journal} {Phys. Rev. A}\ }\textbf {\bibinfo
  {volume} {81}},\ \bibinfo {pages} {053622} (\bibinfo {year}
  {2010})}\BibitemShut {NoStop}%
\bibitem [{\citenamefont {Blankenbecler}\ \emph {et~al.}(1981)\citenamefont
  {Blankenbecler}, \citenamefont {Scalapino},\ and\ \citenamefont
  {Sugar}}]{Blankenbecler:1981vj}%
  \BibitemOpen
  \bibfield  {author} {\bibinfo {author} {\bibfnamefont {R.}~\bibnamefont
  {Blankenbecler}}, \bibinfo {author} {\bibfnamefont {D.~J.}\ \bibnamefont
  {Scalapino}}, \ and\ \bibinfo {author} {\bibfnamefont {R.~L.}\ \bibnamefont
  {Sugar}},\ }\href {http://prd.aps.org/abstract/PRD/v24/i8/p2278_1} {\bibfield
   {journal} {\bibinfo  {journal} {Phys. Rev. D}\ }\textbf {\bibinfo {volume}
  {24}},\ \bibinfo {pages} {2278} (\bibinfo {year} {1981})}\BibitemShut
  {NoStop}%
\bibitem [{\citenamefont {Wu}\ and\ \citenamefont {Zhang}(2005)}]{Wu:2005im}%
  \BibitemOpen
  \bibfield  {author} {\bibinfo {author} {\bibfnamefont {C.}~\bibnamefont
  {Wu}}\ and\ \bibinfo {author} {\bibfnamefont {S.-C.}\ \bibnamefont {Zhang}},\
  }\href {\doibase 10.1103/PhysRevB.71.155115} {\bibfield  {journal} {\bibinfo
  {journal} {Phys. Rev. B}\ }\textbf {\bibinfo {volume} {71}},\ \bibinfo
  {pages} {155115} (\bibinfo {year} {2005})}\BibitemShut {NoStop}%
\bibitem [{\citenamefont {Rubtsov}\ \emph {et~al.}(2005)\citenamefont
  {Rubtsov}, \citenamefont {Savkin},\ and\ \citenamefont
  {Lichtenstein}}]{Rubtsov:2005iw}%
  \BibitemOpen
  \bibfield  {author} {\bibinfo {author} {\bibfnamefont {A.}~\bibnamefont
  {Rubtsov}}, \bibinfo {author} {\bibfnamefont {V.}~\bibnamefont {Savkin}}, \
  and\ \bibinfo {author} {\bibfnamefont {A.}~\bibnamefont {Lichtenstein}},\
  }\href {http://link.aps.org/doi/10.1103/PhysRevB.72.035122} {\bibfield
  {journal} {\bibinfo  {journal} {Phys. Rev. B}\ }\textbf {\bibinfo {volume}
  {72}},\ \bibinfo {pages} {035122} (\bibinfo {year} {2005})}\BibitemShut
  {NoStop}%
\bibitem [{\citenamefont {Greif}\ \emph {et~al.}(2013)\citenamefont {Greif},
  \citenamefont {Uehlinger}, \citenamefont {Jotzu}, \citenamefont {Tarruell},\
  and\ \citenamefont {Esslinger}}]{Greif:2013kb}%
  \BibitemOpen
  \bibfield  {author} {\bibinfo {author} {\bibfnamefont {D.}~\bibnamefont
  {Greif}}, \bibinfo {author} {\bibfnamefont {T.}~\bibnamefont {Uehlinger}},
  \bibinfo {author} {\bibfnamefont {G.}~\bibnamefont {Jotzu}}, \bibinfo
  {author} {\bibfnamefont {L.}~\bibnamefont {Tarruell}}, \ and\ \bibinfo
  {author} {\bibfnamefont {T.}~\bibnamefont {Esslinger}},\ }\href
  {http://www.sciencemag.org/cgi/doi/10.1126/science.1236362} {\bibfield
  {journal} {\bibinfo  {journal} {Science}\ }\textbf {\bibinfo {volume}
  {340}},\ \bibinfo {pages} {1307} (\bibinfo {year} {2013})}\BibitemShut
  {NoStop}%
\bibitem [{\citenamefont {Gull}\ \emph {et~al.}(2011)\citenamefont {Gull},
  \citenamefont {Millis}, \citenamefont {Lichtenstein}, \citenamefont
  {Rubtsov}, \citenamefont {Troyer},\ and\ \citenamefont
  {Werner}}]{Gull:2011jd}%
  \BibitemOpen
  \bibfield  {author} {\bibinfo {author} {\bibfnamefont {E.}~\bibnamefont
  {Gull}}, \bibinfo {author} {\bibfnamefont {A.~J.}\ \bibnamefont {Millis}},
  \bibinfo {author} {\bibfnamefont {A.~I.}\ \bibnamefont {Lichtenstein}},
  \bibinfo {author} {\bibfnamefont {A.~N.}\ \bibnamefont {Rubtsov}}, \bibinfo
  {author} {\bibfnamefont {M.}~\bibnamefont {Troyer}}, \ and\ \bibinfo {author}
  {\bibfnamefont {P.}~\bibnamefont {Werner}},\ }\href
  {http://link.aps.org/doi/10.1103/RevModPhys.83.349} {\bibfield  {journal}
  {\bibinfo  {journal} {Rev. Mod. Phys.}\ }\textbf {\bibinfo {volume} {83}},\
  \bibinfo {pages} {349} (\bibinfo {year} {2011})}\BibitemShut {NoStop}%
\bibitem [{\citenamefont {Kozik}\ \emph {et~al.}(2013)\citenamefont {Kozik},
  \citenamefont {Burovski}, \citenamefont {Scarola},\ and\ \citenamefont
  {Troyer}}]{Kozik:2013ji}%
  \BibitemOpen
  \bibfield  {author} {\bibinfo {author} {\bibfnamefont {E.}~\bibnamefont
  {Kozik}}, \bibinfo {author} {\bibfnamefont {E.}~\bibnamefont {Burovski}},
  \bibinfo {author} {\bibfnamefont {V.~W.}\ \bibnamefont {Scarola}}, \ and\
  \bibinfo {author} {\bibfnamefont {M.}~\bibnamefont {Troyer}},\ }\href
  {http://link.aps.org/doi/10.1103/PhysRevB.87.205102} {\bibfield  {journal}
  {\bibinfo  {journal} {Phys. Rev. B}\ }\textbf {\bibinfo {volume} {87}},\
  \bibinfo {pages} {205102} (\bibinfo {year} {2013})}\BibitemShut {NoStop}%
\bibitem [{\citenamefont {Nomura}\ \emph {et~al.}(2014)\citenamefont {Nomura},
  \citenamefont {Sakai},\ and\ \citenamefont {Arita}}]{Anonymous:2014by}%
  \BibitemOpen
  \bibfield  {author} {\bibinfo {author} {\bibfnamefont {Y.}~\bibnamefont
  {Nomura}}, \bibinfo {author} {\bibfnamefont {S.}~\bibnamefont {Sakai}}, \
  and\ \bibinfo {author} {\bibfnamefont {R.}~\bibnamefont {Arita}},\ }\href
  {\doibase 10.1103/PhysRevB.89.195146} {\bibfield  {journal} {\bibinfo
  {journal} {Phys. Rev. B}\ }\textbf {\bibinfo {volume} {89}},\ \bibinfo
  {pages} {195146} (\bibinfo {year} {2014})}\BibitemShut {NoStop}%
\bibitem [{Note2()}]{Note2}%
  \BibitemOpen
  \bibinfo {note} {When $t_{\delimiter "3222378 }=t_{\delimiter "3223379 }$,
  one could use a shift tuned slightly away from $\Gamma =-U/4$ to gain finite
  weight for odd expansion orders,\cite {Rubtsov:2005iw, Assaad:2007be} thus
  avoiding the inconvenience of correlated double-vertex updates. However, for
  general asymmetric $t_{\delimiter "3222378 }\not =t_{\delimiter "3223379 }$
  case this leads to sign problem in the CT-INT simulation.}\BibitemShut
  {Stop}%
\bibitem [{\citenamefont {Wang}\ \emph
  {et~al.}(2015{\natexlab{c}})\citenamefont {Wang}, \citenamefont {Liu},
  \citenamefont {Imri{\v s}ka}, \citenamefont {Ma},\ and\ \citenamefont
  {Troyer}}]{Wang:2015bva}%
  \BibitemOpen
  \bibfield  {author} {\bibinfo {author} {\bibfnamefont {L.}~\bibnamefont
  {Wang}}, \bibinfo {author} {\bibfnamefont {Y.-H.}\ \bibnamefont {Liu}},
  \bibinfo {author} {\bibfnamefont {J.}~\bibnamefont {Imri{\v s}ka}}, \bibinfo
  {author} {\bibfnamefont {P.~N.}\ \bibnamefont {Ma}}, \ and\ \bibinfo {author}
  {\bibfnamefont {M.}~\bibnamefont {Troyer}},\ }\href {\doibase
  10.1103/PhysRevX.5.031007} {\bibfield  {journal} {\bibinfo  {journal} {Phys.
  Rev. X}\ }\textbf {\bibinfo {volume} {5}},\ \bibinfo {pages} {031007}
  (\bibinfo {year} {2015}{\natexlab{c}})}\BibitemShut {NoStop}%
\bibitem [{Note3()}]{Note3}%
  \BibitemOpen
  \bibinfo {note} {$n$ is a constant that only depends on the lattice geometry
  and the value of $\Delta _R$.}\BibitemShut {Stop}%
\bibitem [{Note4()}]{Note4}%
  \BibitemOpen
  \bibinfo {note} {Long distance shift will be rejected with high probability.
  We therefore also impose the ``closeness'' condition and only shift the
  vertices within the cutoffs $\Delta _{\tau }$ and $\Delta _{R}$.}\BibitemShut
  {Stop}%
\bibitem [{\citenamefont {Rombouts}\ \emph {et~al.}(1999)\citenamefont
  {Rombouts}, \citenamefont {Heyde},\ and\ \citenamefont
  {Jachowicz}}]{1999PhRvL..82.4155R}%
  \BibitemOpen
  \bibfield  {author} {\bibinfo {author} {\bibfnamefont {S.~M.~A.}\
  \bibnamefont {Rombouts}}, \bibinfo {author} {\bibfnamefont {K.}~\bibnamefont
  {Heyde}}, \ and\ \bibinfo {author} {\bibfnamefont {N.}~\bibnamefont
  {Jachowicz}},\ }\href
  {http://adsabs.harvard.edu/cgi-bin/nph-data_query?bibcode=1999PhRvL..82.4155R&link_type=ABSTRACT}
  {\bibfield  {journal} {\bibinfo  {journal} {Phys. Rev. Lett.}\ }\textbf
  {\bibinfo {volume} {82}},\ \bibinfo {pages} {4155} (\bibinfo {year}
  {1999})}\BibitemShut {NoStop}%
\bibitem [{\citenamefont {Loh~Jr}\ and\ \citenamefont
  {Gubernatis}(1992)}]{Loh:1992}%
  \BibitemOpen
  \bibfield  {author} {\bibinfo {author} {\bibfnamefont {E.~Y.}\ \bibnamefont
  {Loh~Jr}}\ and\ \bibinfo {author} {\bibfnamefont {J.~E.}\ \bibnamefont
  {Gubernatis}},\ }\href {http://quest.ucdavis.edu/tutorial/qmc_article.pdf}
  {\emph {\bibinfo {title} {{Electronic Phase Transitions}}}}\ (\bibinfo
  {publisher} {Elsevier Science Publishers},\ \bibinfo {year} {1992})\ p.\
  \bibinfo {pages} {177}\BibitemShut {NoStop}%
\bibitem [{\citenamefont {Woodbury}(1950)}]{woodbury1950inverting}%
  \BibitemOpen
  \bibfield  {author} {\bibinfo {author} {\bibfnamefont {M.~A.}\ \bibnamefont
  {Woodbury}},\ }\href {https://en.wikipedia.org/wiki/Woodbury_matrix_identity}
  {\bibfield  {journal} {\bibinfo  {journal} {Princeton University Statistical
  Research Group, Memo. Rep}\ }\textbf {\bibinfo {volume} {42}} (\bibinfo
  {year} {1950})}\BibitemShut {NoStop}%
\bibitem [{Note5()}]{Note5}%
  \BibitemOpen
  \bibinfo {note} {They are calculated at a random imaginary time during the
  sweep.}\BibitemShut {Stop}%
\bibitem [{\citenamefont {Mermin}\ and\ \citenamefont
  {Wagner}(1966)}]{PhysRevLett.17.1133}%
  \BibitemOpen
  \bibfield  {author} {\bibinfo {author} {\bibfnamefont {N.~D.}\ \bibnamefont
  {Mermin}}\ and\ \bibinfo {author} {\bibfnamefont {H.}~\bibnamefont
  {Wagner}},\ }\href {\doibase 10.1103/PhysRevLett.17.1133} {\bibfield
  {journal} {\bibinfo  {journal} {Phys. Rev. Lett.}\ }\textbf {\bibinfo
  {volume} {17}},\ \bibinfo {pages} {1133} (\bibinfo {year}
  {1966})}\BibitemShut {NoStop}%
\bibitem [{\citenamefont {Salas}\ and\ \citenamefont
  {Sokal}(2000)}]{Salas:1999qh}%
  \BibitemOpen
  \bibfield  {author} {\bibinfo {author} {\bibfnamefont {J.}~\bibnamefont
  {Salas}}\ and\ \bibinfo {author} {\bibfnamefont {A.~D.}\ \bibnamefont
  {Sokal}},\ }\href
  {http://link.springer.com/article/10.1023%2FA%3A1018611122166} {\bibfield
  {journal} {\bibinfo  {journal} {Journal of Statistical Physics}\ }\textbf
  {\bibinfo {volume} {98}},\ \bibinfo {pages} {551} (\bibinfo {year}
  {2000})}\BibitemShut {NoStop}%
\bibitem [{\citenamefont {Sorella}\ and\ \citenamefont
  {Tosatti}(1992)}]{Sorella:1992wd}%
  \BibitemOpen
  \bibfield  {author} {\bibinfo {author} {\bibfnamefont {S.}~\bibnamefont
  {Sorella}}\ and\ \bibinfo {author} {\bibfnamefont {E.}~\bibnamefont
  {Tosatti}},\ }\href {http://iopscience.iop.org/0295-5075/19/8/007} {\bibfield
   {journal} {\bibinfo  {journal} {EPL}\ }\textbf {\bibinfo {volume} {19}},\
  \bibinfo {pages} {699} (\bibinfo {year} {1992})}\BibitemShut {NoStop}%
\bibitem [{\citenamefont {Meng}\ \emph {et~al.}(2010)\citenamefont {Meng},
  \citenamefont {Lang}, \citenamefont {Wessel}, \citenamefont {Assaad},\ and\
  \citenamefont {Muramatsu}}]{Meng:2010gc}%
  \BibitemOpen
  \bibfield  {author} {\bibinfo {author} {\bibfnamefont {Z.~Y.}\ \bibnamefont
  {Meng}}, \bibinfo {author} {\bibfnamefont {T.~C.}\ \bibnamefont {Lang}},
  \bibinfo {author} {\bibfnamefont {S.}~\bibnamefont {Wessel}}, \bibinfo
  {author} {\bibfnamefont {F.~F.}\ \bibnamefont {Assaad}}, \ and\ \bibinfo
  {author} {\bibfnamefont {A.}~\bibnamefont {Muramatsu}},\ }\href
  {http://dx.doi.org/10.1038/nature08942} {\bibfield  {journal} {\bibinfo
  {journal} {Nature}\ }\textbf {\bibinfo {volume} {464}},\ \bibinfo {pages}
  {847} (\bibinfo {year} {2010})}\BibitemShut {NoStop}%
\bibitem [{\citenamefont {Herbut}(2006)}]{Herbut:2006jaa}%
  \BibitemOpen
  \bibfield  {author} {\bibinfo {author} {\bibfnamefont {I.~F.}\ \bibnamefont
  {Herbut}},\ }\href {http://link.aps.org/doi/10.1103/PhysRevLett.97.146401}
  {\bibfield  {journal} {\bibinfo  {journal} {Phys. Rev. Lett.}\ }\textbf
  {\bibinfo {volume} {97}},\ \bibinfo {pages} {146401} (\bibinfo {year}
  {2006})}\BibitemShut {NoStop}%
\bibitem [{\citenamefont {Sorella}\ \emph {et~al.}(2012)\citenamefont
  {Sorella}, \citenamefont {Otsuka},\ and\ \citenamefont
  {Yunoki}}]{Sorella:2012hib}%
  \BibitemOpen
  \bibfield  {author} {\bibinfo {author} {\bibfnamefont {S.}~\bibnamefont
  {Sorella}}, \bibinfo {author} {\bibfnamefont {Y.}~\bibnamefont {Otsuka}}, \
  and\ \bibinfo {author} {\bibfnamefont {S.}~\bibnamefont {Yunoki}},\ }\href
  {http://www.nature.com/doifinder/10.1038/srep00992} {\bibfield  {journal}
  {\bibinfo  {journal} {Sci. Rep.}\ }\textbf {\bibinfo {volume} {2}} (\bibinfo
  {year} {2012})}\BibitemShut {NoStop}%
\bibitem [{\citenamefont {Assaad}\ and\ \citenamefont
  {Herbut}(2013)}]{Assaad:2013kg}%
  \BibitemOpen
  \bibfield  {author} {\bibinfo {author} {\bibfnamefont {F.~F.}\ \bibnamefont
  {Assaad}}\ and\ \bibinfo {author} {\bibfnamefont {I.~F.}\ \bibnamefont
  {Herbut}},\ }\href {http://link.aps.org/doi/10.1103/PhysRevX.3.031010}
  {\bibfield  {journal} {\bibinfo  {journal} {Phys. Rev. X}\ }\textbf {\bibinfo
  {volume} {3}},\ \bibinfo {pages} {031010} (\bibinfo {year}
  {2013})}\BibitemShut {NoStop}%
\bibitem [{\citenamefont {Herbut}\ \emph {et~al.}(2009)\citenamefont {Herbut},
  \citenamefont {Juri{\v c}i{\'c}},\ and\ \citenamefont
  {Vafek}}]{Herbut:2009gaa}%
  \BibitemOpen
  \bibfield  {author} {\bibinfo {author} {\bibfnamefont {I.}~\bibnamefont
  {Herbut}}, \bibinfo {author} {\bibfnamefont {V.}~\bibnamefont {Juri{\v
  c}i{\'c}}}, \ and\ \bibinfo {author} {\bibfnamefont {O.}~\bibnamefont
  {Vafek}},\ }\href {\doibase 10.1103/PhysRevB.80.075432} {\bibfield  {journal}
  {\bibinfo  {journal} {Phys. Rev. B}\ }\textbf {\bibinfo {volume} {80}},\
  \bibinfo {pages} {075432} (\bibinfo {year} {2009})}\BibitemShut {NoStop}%
\bibitem [{Note6()}]{Note6}%
  \BibitemOpen
  \bibinfo {note} {The critical exponents are also different from the ones
  obtained for spinless fermions on honeycomb and $\pi $ flux lattices\cite
  {Wang:2014iba, Wang:2015tf, Li:2015cwb} possibly due to a different number of
  fermion flavors.}\BibitemShut {Stop}%
\bibitem [{\citenamefont {Sandvik}(2012)}]{Sandvik:2012fka}%
  \BibitemOpen
  \bibfield  {author} {\bibinfo {author} {\bibfnamefont {A.~W.}\ \bibnamefont
  {Sandvik}},\ }\href {\doibase 10.1103/PhysRevB.85.134407} {\bibfield
  {journal} {\bibinfo  {journal} {Phys. Rev. B}\ }\textbf {\bibinfo {volume}
  {85}},\ \bibinfo {pages} {134407} (\bibinfo {year} {2012})}\BibitemShut
  {NoStop}%
\bibitem [{\citenamefont {Parisen~Toldin}\ \emph {et~al.}(2015)\citenamefont
  {Parisen~Toldin}, \citenamefont {Hohenadler}, \citenamefont {Assaad},\ and\
  \citenamefont {Herbut}}]{ParisenToldin:2015gs}%
  \BibitemOpen
  \bibfield  {author} {\bibinfo {author} {\bibfnamefont {F.}~\bibnamefont
  {Parisen~Toldin}}, \bibinfo {author} {\bibfnamefont {M.}~\bibnamefont
  {Hohenadler}}, \bibinfo {author} {\bibfnamefont {F.~F.}\ \bibnamefont
  {Assaad}}, \ and\ \bibinfo {author} {\bibfnamefont {I.~F.}\ \bibnamefont
  {Herbut}},\ }\href {\doibase 10.1103/PhysRevB.91.165108} {\bibfield
  {journal} {\bibinfo  {journal} {Phys. Rev. B}\ }\textbf {\bibinfo {volume}
  {91}},\ \bibinfo {pages} {165108} (\bibinfo {year} {2015})}\BibitemShut
  {NoStop}%
\bibitem [{\citenamefont {Hart}\ \emph {et~al.}(2015)\citenamefont {Hart},
  \citenamefont {Duarte}, \citenamefont {Yang}, \citenamefont {Liu},
  \citenamefont {Paiva}, \citenamefont {Khatami}, \citenamefont {Scalettar},
  \citenamefont {Trivedi}, \citenamefont {Huse},\ and\ \citenamefont
  {Hulet}}]{Hart:2015exa}%
  \BibitemOpen
  \bibfield  {author} {\bibinfo {author} {\bibfnamefont {R.~A.}\ \bibnamefont
  {Hart}}, \bibinfo {author} {\bibfnamefont {P.~M.}\ \bibnamefont {Duarte}},
  \bibinfo {author} {\bibfnamefont {T.-L.}\ \bibnamefont {Yang}}, \bibinfo
  {author} {\bibfnamefont {X.}~\bibnamefont {Liu}}, \bibinfo {author}
  {\bibfnamefont {T.}~\bibnamefont {Paiva}}, \bibinfo {author} {\bibfnamefont
  {E.}~\bibnamefont {Khatami}}, \bibinfo {author} {\bibfnamefont {R.~T.}\
  \bibnamefont {Scalettar}}, \bibinfo {author} {\bibfnamefont {N.}~\bibnamefont
  {Trivedi}}, \bibinfo {author} {\bibfnamefont {D.~A.}\ \bibnamefont {Huse}}, \
  and\ \bibinfo {author} {\bibfnamefont {R.~G.}\ \bibnamefont {Hulet}},\ }\href
  {\doibase 10.1038/nature14223} {\bibfield  {journal} {\bibinfo  {journal}
  {Nature}\ }\textbf {\bibinfo {volume} {519}},\ \bibinfo {pages} {211}
  (\bibinfo {year} {2015})}\BibitemShut {NoStop}%
\bibitem [{\citenamefont {Gukelberger}\ \emph {et~al.}(2014)\citenamefont
  {Gukelberger}, \citenamefont {Kozik}, \citenamefont {Pollet}, \citenamefont
  {Prokof'ev}, \citenamefont {Sigrist}, \citenamefont {Svistunov},\ and\
  \citenamefont {Troyer}}]{PhysRevLett.113.195301}%
  \BibitemOpen
  \bibfield  {author} {\bibinfo {author} {\bibfnamefont {J.}~\bibnamefont
  {Gukelberger}}, \bibinfo {author} {\bibfnamefont {E.}~\bibnamefont {Kozik}},
  \bibinfo {author} {\bibfnamefont {L.}~\bibnamefont {Pollet}}, \bibinfo
  {author} {\bibfnamefont {N.}~\bibnamefont {Prokof'ev}}, \bibinfo {author}
  {\bibfnamefont {M.}~\bibnamefont {Sigrist}}, \bibinfo {author} {\bibfnamefont
  {B.}~\bibnamefont {Svistunov}}, \ and\ \bibinfo {author} {\bibfnamefont
  {M.}~\bibnamefont {Troyer}},\ }\href {\doibase
  10.1103/PhysRevLett.113.195301} {\bibfield  {journal} {\bibinfo  {journal}
  {Phys. Rev. Lett.}\ }\textbf {\bibinfo {volume} {113}},\ \bibinfo {pages}
  {195301} (\bibinfo {year} {2014})}\BibitemShut {NoStop}%
\bibitem [{\citenamefont {Shinaoka}\ \emph {et~al.}(2015)\citenamefont
  {Shinaoka}, \citenamefont {Nomura}, \citenamefont {Biermann}, \citenamefont
  {Troyer},\ and\ \citenamefont {Werner}}]{Shinaoka:2015wq}%
  \BibitemOpen
  \bibfield  {author} {\bibinfo {author} {\bibfnamefont {H.}~\bibnamefont
  {Shinaoka}}, \bibinfo {author} {\bibfnamefont {Y.}~\bibnamefont {Nomura}},
  \bibinfo {author} {\bibfnamefont {S.}~\bibnamefont {Biermann}}, \bibinfo
  {author} {\bibfnamefont {M.}~\bibnamefont {Troyer}}, \ and\ \bibinfo {author}
  {\bibfnamefont {P.}~\bibnamefont {Werner}},\ }\href
  {http://arxiv.org/abs/1508.06741v1} {\bibfield  {journal} {\bibinfo
  {journal} {arXiv:1508.06741}\ } (\bibinfo {year} {2015})}\BibitemShut
  {NoStop}%
\bibitem [{\citenamefont {Bauer}\ \emph {et~al.}(2011)\citenamefont {Bauer},
  \citenamefont {Carr}, \citenamefont {Evertz}, \citenamefont {Feiguin},
  \citenamefont {Freire}, \citenamefont {Fuchs}, \citenamefont {Gamper},
  \citenamefont {Gukelberger}, \citenamefont {Gull}, \citenamefont {Guertler},
  \citenamefont {Hehn}, \citenamefont {Igarashi}, \citenamefont {Isakov},
  \citenamefont {Koop}, \citenamefont {Ma}, \citenamefont {Mates},
  \citenamefont {Matsuo}, \citenamefont {Parcollet}, \citenamefont {Pawlowski},
  \citenamefont {Picon}, \citenamefont {Pollet}, \citenamefont {Santos},
  \citenamefont {Scarola}, \citenamefont {Schollwock}, \citenamefont {Silva},
  \citenamefont {Surer}, \citenamefont {Todo}, \citenamefont {Trebst},
  \citenamefont {Troyer}, \citenamefont {Wall}, \citenamefont {Werner},\ and\
  \citenamefont {Wessel}}]{BBauer:2011tz}%
  \BibitemOpen
  \bibfield  {author} {\bibinfo {author} {\bibfnamefont {B.}~\bibnamefont
  {Bauer}}, \bibinfo {author} {\bibfnamefont {L.~D.}\ \bibnamefont {Carr}},
  \bibinfo {author} {\bibfnamefont {H.~G.}\ \bibnamefont {Evertz}}, \bibinfo
  {author} {\bibfnamefont {A.}~\bibnamefont {Feiguin}}, \bibinfo {author}
  {\bibfnamefont {J.}~\bibnamefont {Freire}}, \bibinfo {author} {\bibfnamefont
  {S.}~\bibnamefont {Fuchs}}, \bibinfo {author} {\bibfnamefont
  {L.}~\bibnamefont {Gamper}}, \bibinfo {author} {\bibfnamefont
  {J.}~\bibnamefont {Gukelberger}}, \bibinfo {author} {\bibfnamefont
  {E.}~\bibnamefont {Gull}}, \bibinfo {author} {\bibfnamefont {S.}~\bibnamefont
  {Guertler}}, \bibinfo {author} {\bibfnamefont {A.}~\bibnamefont {Hehn}},
  \bibinfo {author} {\bibfnamefont {R.}~\bibnamefont {Igarashi}}, \bibinfo
  {author} {\bibfnamefont {S.~V.}\ \bibnamefont {Isakov}}, \bibinfo {author}
  {\bibfnamefont {D.}~\bibnamefont {Koop}}, \bibinfo {author} {\bibfnamefont
  {P.~N.}\ \bibnamefont {Ma}}, \bibinfo {author} {\bibfnamefont
  {P.}~\bibnamefont {Mates}}, \bibinfo {author} {\bibfnamefont
  {H.}~\bibnamefont {Matsuo}}, \bibinfo {author} {\bibfnamefont
  {O.}~\bibnamefont {Parcollet}}, \bibinfo {author} {\bibfnamefont
  {G.}~\bibnamefont {Pawlowski}}, \bibinfo {author} {\bibfnamefont {J.~D.}\
  \bibnamefont {Picon}}, \bibinfo {author} {\bibfnamefont {L.}~\bibnamefont
  {Pollet}}, \bibinfo {author} {\bibfnamefont {E.}~\bibnamefont {Santos}},
  \bibinfo {author} {\bibfnamefont {V.~W.}\ \bibnamefont {Scarola}}, \bibinfo
  {author} {\bibfnamefont {U.}~\bibnamefont {Schollwock}}, \bibinfo {author}
  {\bibfnamefont {C.}~\bibnamefont {Silva}}, \bibinfo {author} {\bibfnamefont
  {B.}~\bibnamefont {Surer}}, \bibinfo {author} {\bibfnamefont
  {S.}~\bibnamefont {Todo}}, \bibinfo {author} {\bibfnamefont {S.}~\bibnamefont
  {Trebst}}, \bibinfo {author} {\bibfnamefont {M.}~\bibnamefont {Troyer}},
  \bibinfo {author} {\bibfnamefont {M.~L.}\ \bibnamefont {Wall}}, \bibinfo
  {author} {\bibfnamefont {P.}~\bibnamefont {Werner}}, \ and\ \bibinfo {author}
  {\bibfnamefont {S.}~\bibnamefont {Wessel}},\ }\href
  {http://iopscience.iop.org/1742-5468/2011/05/P05001} {\bibfield  {journal}
  {\bibinfo  {journal} {J. Stat. Mech.: Theor. Exp.}\ }\textbf {\bibinfo
  {volume} {2011}},\ \bibinfo {pages} {P05001} (\bibinfo {year}
  {2011})}\BibitemShut {NoStop}%
\bibitem [{\citenamefont {Assaad}\ and\ \citenamefont
  {Lang}(2007)}]{Assaad:2007be}%
  \BibitemOpen
  \bibfield  {author} {\bibinfo {author} {\bibfnamefont {F.}~\bibnamefont
  {Assaad}}\ and\ \bibinfo {author} {\bibfnamefont {T.}~\bibnamefont {Lang}},\
  }\href {http://link.aps.org/doi/10.1103/PhysRevB.76.035116} {\bibfield
  {journal} {\bibinfo  {journal} {Phys. Rev. B}\ }\textbf {\bibinfo {volume}
  {76}},\ \bibinfo {pages} {035116} (\bibinfo {year} {2007})}\BibitemShut
  {NoStop}%
\end{thebibliography}%
\end{document}